\documentclass[aps,11pt,prd,groupedaddress,nofootinbib,notitlepage,eqsecnum,preprintnumbers]{revtex4-2}
	
\usepackage{subcaption}
\usepackage{ragged2e}
\DeclareCaptionJustification{justified}{\justifying}
\makeatletter
\newcommand{\subsetsim}{\mathrel{\mathpalette\subset@sim\relax}}
\newcommand{\subset@sim}[2]{%
  \vtop{\offinterlineskip\m@th
    \ialign{\hfil##\cr
      $#1\subset$\cr\noalign{\kern0.5pt}\scalebox{0.9}{$#1\sim$}\cr
    }%
  }%
}
\makeatother

\usepackage{amssymb,amsmath,verbatim,mathtools,needspace,enumitem,etoolbox,graphicx,microtype,afterpage,bm}

\usepackage[dvipsnames, usenames]{xcolor}

\definecolor{linkcolor}{rgb}{0.0,0.3,0.5}
\usepackage{booktabs}
\usepackage[unicode, colorlinks=true, linkcolor=linkcolor, citecolor=linkcolor, filecolor=linkcolor,urlcolor=linkcolor, pdfusetitle]{hyperref}
\usepackage[all]{hypcap}
\usepackage[T1]{fontenc}
\usepackage[utf8]{inputenc}
\usepackage{tabularx}
\usepackage{float}
\usepackage{multirow}
\usepackage{pifont}
\usepackage{lmodern}

\usepackage[normalem]{ulem}

\allowdisplaybreaks
\usepackage{tikz}
\usetikzlibrary{shapes.misc, positioning, arrows.meta}
\usepackage{framed}
\usepackage{hyperref}
\hypersetup{colorlinks, citecolor=bluscuro, linkcolor=black, urlcolor=bluscuro}
\definecolor{rossos}{cmyk}{0,1,1,0.55}
\definecolor{bluscuro}{rgb}{0.15, 0.2, .85}
\definecolor{bluchiaro}{cmyk}{1,.3,0.,0.1}
\definecolor{ForestGreen}{rgb}{0.13, 0.55, 0.13}
\definecolor{TLGreen}{RGB}{50, 164, 49}
\definecolor{TLOrange}{RGB}{231,180,22}
\definecolor{TLRed}{RGB}{204,50,50}

\newcommand{\TLBullet}[1]{\raisebox{-5pt}{\scalebox{0.23}{\begin{tikzpicture}\shadedraw[rounded corners=15pt, top color=gray!84!black,bottom color=black, line width=.6pt] (0,0) rectangle ++(6,2); \ifthenelse{#1=1}{\draw[fill=green,line width=1.pt]  (1,1) circle(.75cm);}{\draw[fill=green!35!black,line width=1.pt]  (1,1) circle(.75cm);}\ifthenelse{#1=2}{\draw[fill=yellow,line width=1.pt]  (3,1) circle(.75cm);}{\draw[fill=yellow!60!black,line width=1.pt]  (3,1) circle(.75cm);}\ifthenelse{#1=3}{\draw[fill=red,line width=1.pt]  (5,1) circle(.75cm);}{\draw[fill=red!50!black,line width=1.pt]  (5,1) circle(.75cm);}\end{tikzpicture}}}}

\newcommand{\bs}{\begin{subequations}}
\newcommand{\es}{\end{subequations}}

\newcommand{\be}{\begin{equation}}
\newcommand{\ee}{\end{equation}}

\def\lsim{\mathrel{\rlap{\lower4pt\hbox{\hskip0.5pt$\sim$}}
    \raise1pt\hbox{$<$}}}         
\def\gsim{\mathrel{\rlap{\lower4pt\hbox{\hskip0.5pt$\sim$}}
    \raise1pt\hbox{$>$}}}         

\usepackage{siunitx}
\DeclareSIUnit \parsec {pc}
\DeclareSIUnit \arcsecondfull {arcsec}
\DeclareSIUnit \year{yr}
\DeclareSIUnit \day{day}
\DeclareSIUnit \hour{hr}
\DeclareSIUnit \radiant{rad}
\DeclareSIUnit \degfull{deg}
\DeclareSIUnit \erg {erg}
\DeclareSIUnit \Lsun {L_\odot}
\DeclareSIUnit \Msun {M_\odot}
\DeclareSIUnit \AstroUnit {au}

\usepackage{nicefrac}

\graphicspath{{./figures/}}

\begin{document}

\title{Probing modified Hawking evaporation with gravitational waves from the primordial black hole dominated universe}

\author{\textsc{Shyam Balaji$^{a}$}}
    \email{{shyam.balaji}@{kcl.ac.uk}}
\author{\textsc{Guillem Dom\`enech$^{b,c}$}}
    \email{{guillem.domenech}@{itp.uni-hannover.de}}
\author{\textsc{Gabriele Franciolini$^{d}$}}
\email{{gabriele.franciolini}@{cern.ch}}
\author{\textsc{Alexander Ganz$^{b}$}}
\email{{alexander.ganz}@{itp.uni-hannover.de}}
\author{\textsc{Jan Tränkle$^{b}$}}
\email{{jan.traenkle}@{itp.uni-hannover.de}}

\affiliation{$^a$ Physics Department, King’s College London, Strand, London, WC2R 2LS, United Kingdom }
\affiliation{$^b$Institute for Theoretical Physics, Leibniz University Hannover, Appelstraße 2, 30167 Hannover, Germany.}
\affiliation{$^c$ Max-Planck-Institut für Gravitationsphysik, Albert-Einstein-Institut, 30167 Hannover, Germany}
\affiliation{$^d$ CERN, Theoretical Physics Department, Esplanade des Particules 1, Geneva 1211, Switzerland}

\date{\today}

\begin{abstract}
It has been recently proposed that Hawking evaporation might slow down after a black hole has lost about half of its mass. Such an effect, called “memory burden”, is parameterized as a suppression in the mass loss rate by negative powers $n$
of the black hole entropy and could considerably extend the lifetime of a black hole. We study the impact of memory burden on the Primordial Black Hole (PBH) reheating scenario. 
Modified PBH evaporation leads to a significantly longer PBH dominated stage. Requiring that PBHs evaporate prior enough to Big Bang Nucleosynthesis shrinks the allowed PBH mass range. Indeed, we find that for $n>2.5$ the PBH reheating scenario is not viable. The frequency of the Gravitational Waves (GWs) induced by PBH number density fluctuations is bound to be larger than about a Hz, while the amplitude of the GW spectrum is enhanced due to the longer PBH dominated phase. Interestingly, we show that, in some models, the slope of the induced GW spectrum might be sensitive to the modifications to Hawking evaporation, proving it may be possible to test the “memory burden” effect via induced GWs. Lastly, we argue that our results could also apply to general modifications of Hawking evaporation.
\end{abstract}

\preprint{CERN-TH-2024-037, ET-0090A-24, KCL-2024-53}

\maketitle

\newpage

\section{Introduction\label{sec:intro}}
The idea of black holes originating in the early universe has been explored for more than half a century \citep{Zeldovich1967,Hawking1971,Carr1974}, with \citet{Chapline1975} being the first to propose that primordial black holes (PBHs) could account for all the dark matter in the universe. Since then, the cosmological role of PBHs has been investigated, taking into consideration the possibility they could have masses that range from $M\sim {\cal O}(10^4) M_{\rm pl}$, with $M_{\rm pl}$ being the Planck mass, up to the “incredulity limit” above $M\sim10^{10}\,M_{\odot}$, where $M_\odot$ is a solar mass. See e.g.\@ Refs.~\cite{Khlopov:2008qy,Sasaki:2018dmp,Carr:2020gox,Green:2020jor,Escriva:2022duf,Ozsoy:2023ryl} for recent reviews. This has resulted in strong limits that rule out PBHs from making up the whole dark matter except for a relatively narrow mass window in the asteroid range $M\in[10^{17},10^{22}]\,\rm{g}$. 

The lower limit of $M\sim 10^{17}\rm g$ is due to constraints from black hole evaporation. This phenomenon was first suggested by \citet{Hawking1974} when examining the phenomenology of light PBHs. He demonstrated that a black hole radiates a thermal spectrum of particles, with the temperature of the radiation proportional to $T\sim 1/M_{\rm PBH}$. Based on the semi-classical computation, the evaporation process is self-similar and terminates with a final explosion as $M\rightarrow0$. 
While a population of light PBHs close to the $M\sim 10^{17}\rm g$ could be detected by searching for the particles emitted in the late-time universe through the evaporation process, even lighter PBHs would evaporate in the early universe, making them very hard to discover.

However, it has been recently argued that some of the conclusions based on black hole evaporation can be evaded. Ref.~\cite{deFreitasPacheco:2023hpb} (see also \cite{Dvali:2021ofp}) examined “quasi-extremal” PBHs, for which the evaporation efficiency is reduced, and showed that a longer PBH lifetime (and associated smaller luminosity) makes them evade bounds and potentially remain a feasible dark matter candidate. Refs.~\cite{Extradims} and \cite{Extradims2, Anchordoqui:2024dxu} have explored PBHs in the framework of large extra dimensions \citep{ADD} and demonstrated that this allows for new mass windows for light PBHs as dark matter candidates.

For 4-dimensional black holes, Ref.~\cite{Dvali2020} argued that the latest instant Hawking's semi-classical calculations can break down is when the black hole has shed about half of its initial mass. Their main argument is that Hawking's result completely disregards the backreaction of the emission on the black hole itself. Refs.~\cite{Dvali2018,Dvali2020} then suggest that this effect should no longer be negligible when the energy of the emitted quanta becomes comparable to that of the black hole itself. The proposed backreaction leads to a “memory burden” \cite{Dvali2018}, which effectively suppresses the evaporation rate by inverse powers of the black hole entropy $S^{-n}$.
This dependence of the suppression on the black hole entropy was conjectured in Ref.~\cite{Dvali2020} and is motivated by the consideration that a black hole, viewed as a quantum system with maximal memory storage capacity, carries a high load of quantum information. When the black hole emits a quantum of Hawking radiation, the information stored previously in a set of gapless “memory” modes has to be re-written to the remaining internal degrees of freedom, which over time becomes increasingly costly in energy as fewer such gapless modes are available. Thus, naively one expects that the suppression of the evaporation rate should be related to the amount of information stored within the black hole, which is exactly what the Bekenstein-Hawking entropy $S$ measures. The free parameter $n>0$ quantifies the precise strength of the suppression.
Since for $M>M_{\rm pl}$ the entropy of a black hole is “large”, the memory burden can significantly extend the black hole’s lifetime. 
Interestingly, Refs.~\cite{Dvali:2021byy, Alexandre:2024nuo, Thoss:2024hsr} recently noted that, in the presence of the memory burden, PBHs with masses below $10^{10}\rm g$ could still account for the whole dark matter.

In this work, we focus on even lighter PBHs which may have dominated the energy density of the universe in its very early stages and subsequently led to reheating via the emission of thermal radiation through Hawking evaporation \cite{Carr:1976zz,Chapline:1976au, Garcia-Bellido:1996mdl, Lidsey:2001nj,Anantua:2008am,Hidalgo:2011fj,RiajulHaque:2023cqe}. 
Crucially, as shown in Refs.~\cite{Inomata:2020lmk,Papanikolaou:2020qtd,Domenech:2020ssp,Domenech:2021wkk,Papanikolaou:2021uhe,Papanikolaou:2022chm,Pearce:2023kxp,Bhaumik:2022pil,Bhaumik:2022zdd} (see also Refs.~\cite{Domenech:2021ztg,Domenech:2023jve} for reviews), a PBH dominated stage in the very early universe could lead to an observable GW background. 
This signature would result from perturbations induced by PBH shot noise at formation, while second-order induced GWs associated with the PBH formation mechanism would fall at frequencies above the LIGO/Virgo/Kagra band \cite{Franciolini:2023osw} in the so-called ultra-high frequency range \cite{Aggarwal:2020olq,Franciolini:2022htd}.
Since the GW production in this scenario depends on the dynamics before and after PBH evaporation, it may represent a unique opportunity to test modifications of Hawking evaporation. In this work, we will show that GWs induced right after evaporation by PBH number density fluctuations have the potential to probe the memory burden model. Also, since the main effect of the memory burden is to extend the duration of the PBH domination without drastically altering the final instants of evaporation, the GW signal is enhanced with respect to the standard case.

This paper is organized as follows. In \S~\ref{sec:PBHdomination} we study the impact of the memory burden on the PBH dominated universe. In \S~\ref{sec:inducedGWs}, we estimate the GW spectrum induced right after evaporation by the PBH number density fluctuations in the extended PBH dominated universe. We end with some discussions in \S~\ref{sec:conclusions}. We provide details of the calculations in the appendix. In particular, in App.~\ref{App:detailedformulas}, we present the detailed formulas of the extended PBH dominated universe, and in App.~\ref{App:evolution}, we give the details on the evolution of the PBH number density fluctuations until PBH evaporation. In App.~\ref{App:iGWspectrum}, the computation of the induced GWs is explained.
We work in natural units where $\hbar=c=1$.

\section{Memory burden and the PBH dominated universe\label{sec:PBHdomination}}

We start by describing the impact of the memory burden on PBH evaporation, and later we turn to study its effects on the PBH dominated universe. Let us assume, as in Refs.~\cite{Alexandre:2024nuo, Thoss:2024hsr}, that the black hole mass loss rate due to Hawking evaporation is given by
\begin{align}\label{eq:dMPBHscdt}
\frac{dM}{dt}=-\frac{A\,M_{\rm pl}^4}{M^2}\times \left\{
\begin{aligned}
&1 &({\rm for}\,\,M\geq M_{\rm mb})\\
&{S^{-n}({\cal M})} &({\rm for}\,\,M<M_{\rm mb})
\end{aligned}
\right.\,,
\end{align}
where 
$M_{\rm mb}=qM_{\rm f}$ parametrizes the moment when the modification to the semi-classical approximation sets in, with $M_{\rm f}$ being the PBH mass at formation and $q$ a free parameter.
The semi-classical (sc) evaporation is recovered by taking the limit $n\to 0$. Note that from now on we use the subscript “f” to denote evaluation at the time of formation.
We introduced ${\cal M}$ to distinguish two different possibilities discussed in Ref.~\cite{Thoss:2024hsr}:
\begin{enumerate}
    \item The entropy in Eq.~\eqref{eq:dMPBHscdt} is a constant factor with ${\cal M}=M_{\rm mb}=qM_{\rm f}$. This effect basically extends the PBH lifetime but does not significantly modify the final evaporation. We will refer to this case as “memory burden 1” (mb1). 
    \item When the entropy in \eqref{eq:dMPBHscdt} is mass dependent, that is ${\cal M}=M$. The effect on the lifetime of the PBH is similar to mb1, but the final instants of evaporation are also modified. We will call this case “memory burden 2” (mb2). 
\end{enumerate}
In Eq.~\eqref{eq:dMPBHscdt}, we also defined the black hole entropy
\begin{align}\label{eq:entropy}
S({\cal M})=\frac{1}{2}\frac{{\cal M}^2}{M_{\rm pl}^2}\approx e^{24}\left(\frac{{\cal M}}{1\,\rm g}\right)^2\,,
\end{align}
where we introduced the $e^{24}$ factor for later compactness of equations, and \cite{Inomata:2020lmk,Hooper:2019gtx,Masina:2020xhk}
\begin{align}
A=\frac{3.8\pi g_H(T_{\rm PBH})}{480}\,,
\label{eq:A}
\end{align}
where $g_H(T_{\rm PBH})$ are the spin-weighted degrees of freedom heavier than the PBH temperature $T_{\rm PBH}=M_{\rm pl}^2/M$. In what follows, we only consider ultra-light PBHs with $M \ll 10^{11}\,{\rm g}$ and, therefore, we use the high temperature SM relativistic degrees of freedom $g_H(T_{\rm PBH})\approx 108$, unless stated otherwise.
It should be noted at this point that the black hole entropy \eqref{eq:entropy} is generally determined by the surface area $A_h$ of the event horizon as ${S=A_h/(4 G_N)}$. In the case of a rotating and/or charged black hole, $A_h$ and, therefore, also $S$ depend on the charge $Q$ and the angular momentum $J$ in addition to the mass $M$. For simplicity, we will focus on non-rotating and non-charged PBHs in this work.

Integrating Eq.~\eqref{eq:dMPBHscdt} we find that
\begin{align}\label{eq:MPBHmb}
M(t)\approx \left\{
\begin{aligned}
&M_{\rm f} \left( 1 - \frac{t}{t_{\rm eva}^{\rm sc}} \right)^{1/3} &({\rm for}\,\,t<t_{\rm mb})\\
&M_{\rm mb}
   \left(1-\frac{t}{t_{\rm eva}}\right)^{{1}/{(3\alpha)}}&({\rm for}\,\,t>t_{\rm mb})
\end{aligned}
\right.\,,
\end{align}
where we defined
\begin{align}\label{eq:tevamb}
t_{\rm eva}\approx \alpha^{-1}q^{3+2
   n} S_{\rm f}^n\, t_{\rm eva}^{\rm sc}~\quad{\rm where}\quad t_{\rm eva}^{\rm sc} = \frac{M_{\rm f}^3}{3 A M_{\rm pl}^4}\approx 4.1\times 10^{-28}\,{\rm s}\left(\frac{M_{\rm f}}{1\,\rm g}\right)^3,
\end{align}
and we introduced the parameter $\alpha$ to distinguish between the mb1 and mb2 cases, namely
\begin{align}
\alpha=\left\{
\begin{aligned}
&1 &{\rm for\,\,mb1}\\
&1+2n/3  &{\rm for\,\,mb2}
\end{aligned}
\right.\,.
\end{align}
By continuity, one has that
\begin{align}\label{eq:tmb}
    t_{\rm mb} = (1-q^3) t^{\rm sc}_{\rm eva}\,.
\end{align}
In Eq.~\eqref{eq:MPBHmb}, we assumed that $n\neq 0$ and $M_{\rm f}\gg M_{\rm pl}$ but we present the explicit expressions of the coefficients in App.~\ref{App:detailedformulas}. Nevertheless, in the present form, the standard case can be recovered by taking the limit $n\to 0$ and $q\to 1$. We also assumed that the time of formation $t_{\rm f}$ is much earlier than the time of evaporation $t_{\rm eva}$. This is always the case for the mass ranges considered. It is also interesting to note that $t_{\rm mb}$ \eqref{eq:tmb} is very close to the evaporation time in the standard case. This is because the timescale to emit a sizeable fraction of PBH mass is comparable to the evaporation time due to the strong scaling of the emission rate with mass. 

We note that our results are also applicable to the scenarios considered in Ref.~\cite{deFreitasPacheco:2023hpb}.
In particular, we can underline a few relevant examples. 
We introduce $\varepsilon$ which parametrizes, in their notation, the departure from the standard Hawking evaporation of a Schwarzschild black hole leading to a temperature $T \sim \frac{[\varepsilon / (1+\varepsilon)^2 ]}{M_{\rm PBH}} $. 
In the case of a nearly extremal Reissner-Nordstrom charged black hole, $\varepsilon = \sqrt{1-Q^2/M^2} \ll 1$, where $Q$ represents the PBH charge.  
To consider these scenarios, it is sufficient to assume $q = 1$ while identifying the factor $\alpha^{-1}q^{3+2n}S_{\rm f}^n$ with $\frac{(1+\varepsilon)^6}{2^6\varepsilon^4}$.
Another notable example is provided by higher dimensional black holes \cite{Extradims,Extradims2}.
An initial BH of size $r_s$ much larger than the size of extra dimensions $R$ evaporates following the 4D decay equation until its horizon becomes $r_s \simeq R$. 
Afterward, it would continue its decay at a rate that is slower than 4D due to the effect of the extra dimension. 
We can derive the corresponding phenomenological quantities by relating Eq.~\eqref{eq:MPBHmb} to Eq.~(A22) in \cite{deFreitasPacheco:2023hpb}.
As a final example, also a large number of additional degrees of freedom beyond the Standard Model could lead to a modified evaporation rate at small PBH masses, which we can describe within our setup by replacing Eq.~\eqref{eq:dMPBHscdt} with Eq.~(14) from \cite{Wang:2023wsm}.

\subsection{Extended PBH reheating}\label{sec:PBHreheating}
\begin{figure}
    \centering
    \includegraphics[width=0.75\textwidth]{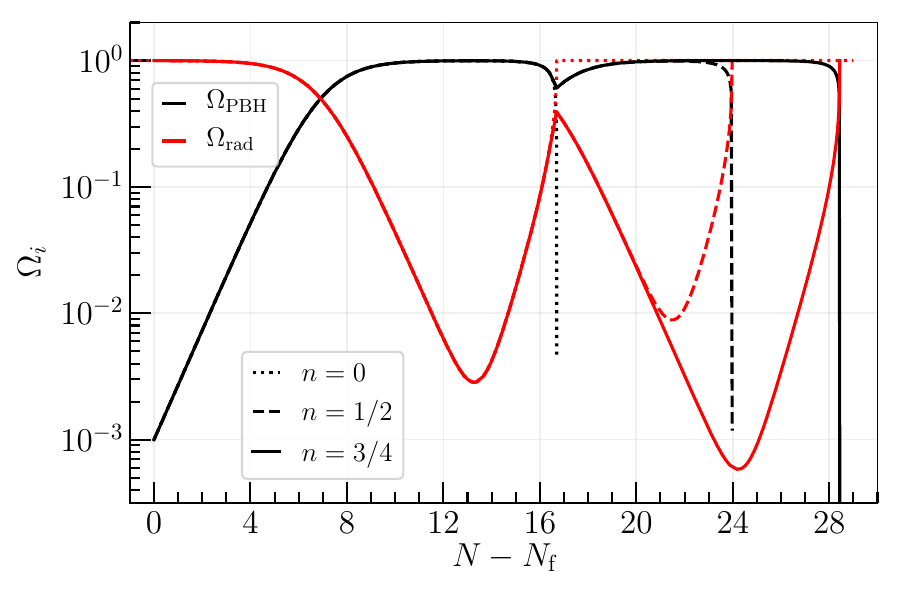}
    \caption{Evolution of the fractional energy densities $\Omega_{\rm PBH}$ (black) and $\Omega_{\rm rad}$ (red) in terms of the number of e-folds $N=\ln a$ since PBH formation at $N_{\rm f}$. The semi-classical evaporation proceeds until $N_{\rm eva}^{\rm sc}\approx 16.7$, and the memory burden effect sets in at $N_{\rm mb}\approx 16.62$. 
For $n=0$, memory burden is switched off, and PBH domination finishes soon after $t_{\rm mb}$ (dotted line).     
For larger values of $n>0$, the PBH evaporation is delayed considerably, and PBH domination lasts longer (dashed and solid lines).
Parameter values for the plot are $M_{\rm f}=10 \si{\gram} ,\ \beta=10^{-3}, \ q=\frac{1}{2}$ and we consider the case mb2.}
    \label{fig:OmegaPBHRadPlot}
\end{figure}

Let us now turn to the PBH reheating scenario (see e.g.\@ Refs.~\cite{Inomata:2020lmk,Domenech:2020ssp,Domenech:2021wkk} for more details on the standard case). We assume that PBHs form from the collapse of large primordial fluctuations after inflation in the radiation dominated universe \cite{Khlopov:2008qy,Sasaki:2018dmp,Carr:2020gox,Green:2020jor,Escriva:2022duf,Ozsoy:2023ryl}.\footnote{Note that PBHs could also form from first order phase transitions \cite{Crawford:1982yz,Kodama:1982sf,Flores:2024lng}, the collapse of  Q-balls \cite{Cotner:2016cvr,Cotner:2019ykd,Flores:2021jas}, fifth forces \cite{Amendola:2017xhl,Flores:2020drq,Domenech:2023afs} and preheating \cite{Martin:2020fgl}.} For analytical simplicity, we further assume a monochromatic PBH mass spectrum (for a broad mass spectrum in the PBH reheating scenario, see Ref.~\cite{Papanikolaou:2022chm}) with the PBH mass given by
\begin{align}
    M_{\rm f} = \frac{4 \pi \gamma M_{\rm pl}^2}{H_{\rm f}}~,
\end{align}
where $\gamma \sim 0.2$ during radiation domination \cite{1975ApJ...201....1C} and $H_{\rm f}$ is the Hubble parameter at the time of formation of the PBHs.
Notice that the Hubble parameter is bounded to be below  $H_{\rm f} < 6 \cdot 10^{13}$GeV from  Planck 2018 and the BICEP2/Keck Array BK14 tight upper bound on the tensor-to-scalar ratio \cite{Planck:2018jri}, leading to a minimum mass $M_{\rm f} \gtrsim 10^4 M_{\rm pl}$.
We also consider that we have an initial fraction $\beta$ of PBHs. Then, the \textit{mean} energy density of PBHs is given by $\rho_{\rm PBH}(t)=M(t)\,n_{\rm PBH}$ where $n_{\rm PBH}$ is the number density of PBHs. By number density conservation, we have that $n_{\rm PBH}\propto a^{-3}$, where $a$ is the scale factor of a Friedmann–Lemaître–Robertson–Walker (FLRW) universe. The initial fraction $\beta$ is related to the energy density of PBHs at formation by
\begin{align}
    \beta =\Omega_{\rm PBH,f}= \frac{\rho_{\rm PBH, f}}{3 H_{\rm f}^2 M_{\rm pl}^2}\,,
\end{align}
where we took the opportunity to define the energy density ratio $\Omega_{\rm PBH}$, which is shown in Fig.~\ref{fig:OmegaPBHRadPlot}. The same definition applies to the radiation component.

After formation, the energy density of the PBHs decays into radiation due to Hawking evaporation. This evolution can be described as
\begin{align}\label{eq:backgroud}
   & \dot \rho_{\rm PBH} + (3 H + \Gamma) \rho_{\rm PBH} =0~, \\
    & \dot \rho_{\rm rad} + 4 H \rho_{\rm rad} - \Gamma \rho_{\rm PBH}=0~,
\end{align}
where $\rho_{\rm rad}$ is the energy density of radiation, and $\Gamma$ is the decay rate due to the Hawking evaporation, namely
\begin{align}
    \Gamma = - \frac{d \log M}{d t} = \begin{cases}
        \frac{1}{3 (t_{\rm eva}^{\rm sc} - t)}\qquad {\rm for} \quad t \leq t_{\rm mb} \\
        \frac{1}{3 \alpha (t_{\rm eva} - t)}\qquad {\rm for} \quad t > t_{\rm mb}
    \end{cases}~.
\end{align}
For $t \ll t_{\rm eva}$ the decay rate $\Gamma \ll H$ can be neglected in comparison to the Hubble expansion and the relative ratio between the energy density of PBHs and the radiation grows in time as the scale factor $a$. Depending on the initial fraction $\beta$,  PBHs may dominate the universe before complete evaporation. We show in Fig.~\ref{fig:OmegaPBHRadPlot} some concrete examples. Due to the memory burden, the evaporation is suppressed by a negative power of the entropy and, therefore, for $n> 0$, the duration of the PBH domination can be significantly enhanced. It is interesting to note that if $\alpha>1$ the final burst happens faster than in the standard case.

In general, we find two possibilities depending on whether the early PBH-radiation equality (which we denote as “eeq” and which is defined by $\rho_{\rm PBH}(t_{\rm eeq})=\rho_{\rm rad}(t_{\rm eeq})$) happens before or after the memory burden time $t_{\rm mb}$ \eqref{eq:tmb}. In the first case, that is $t_{\rm mb} < t_{\rm eeq}$, the initial fraction of the PBHs is too low to reach matter radiation equality before the memory burden time. In fact, this case roughly corresponds to the standard scenario where PBHs evaporate before dominating the universe. Now, due to the memory burden, PBHs have a second chance to dominate the universe. In the second case, where $t_{\rm mb} > t_{\rm eeq}$, the early PBH-radiation equality occurs before the memory burden time. Then, as seen in Fig.~\ref{fig:OmegaPBHRadPlot}, there is a short period in the PBH dominated stage when the PBH energy density decays, but PBHs still dominate the universe. We find that, in that case, $\Omega_{\rm PBH}(t_{\rm mb})\approx 0.59$ for $q=1/2$.
It should be noted that the precise value of $\Omega_{\rm PBH}(t_{\rm mb})$ depends on $q$. For example, for $q\lesssim0.41$ we find that $\Omega_{\rm PBH}(t_{\rm mb})<1/2$ and radiation temporarily dominates the universe.
Depending on the parameters of the modified evaporation, we find that for small enough values of $q$ PBHs do not come to dominate again after $t_{\rm mb}$, while for other choices there is a second stage of PBH domination. We do not consider these possibilities here and focus on the regime $0.41<q<1$. In particular, we will take $q=\frac{1}{2}$ for all plots, unless stated otherwise.
Also, note that the PBH remnant scenario would correspond to $q\sim M_{\rm pl}/M_{\rm f}$ (and $n\to\infty$ so that Hawking evaporation stops completely).

Since the eeq time depends on the initial PBH fraction $\beta$, we find that the limiting case where $t_{\rm mb} \sim t_{\rm eeq}$ corresponds to
\begin{align}
    \beta_*= & \frac{1}{4}\sqrt{\frac{3A}{\sqrt{2}\pi\gamma(1-q^3)S_{\rm f}}} \simeq \frac{4.6\times 10^{-6}}{\sqrt{1-q^3}}\left(\frac{M_{\rm f}}{1\,\rm g}\right)^{-1}\,.
    \label{eq:lower_bound_beta_standard}
\end{align}
For $\beta<\beta_*$, the eeq time happens after the memory burden time. For $\beta>\beta_*$, the eeq occurs before the memory burden is activated. We find that qualitatively the two possibilities yield very similar results regarding the PBH reheating scenario. The only quantitative difference is a shift in the time of eeq, which amounts to replacing $t_{\rm eeq}\to q^{-2} t_{\rm eeq}$ ($a_{\rm eeq}/a_{\rm f}=q^{-1}\beta^{-1})$ when $\beta<\beta_*$. Assuming $q\sim 1/2$, this only yields a factor ${\cal O}(1-10)$ difference. For this reason, we neglect this difference in our derivations, although we explain how to recover this effect at the appropriate places.

We obtain the ratio between the eeq and the evaporation times \eqref{eq:tevamb} as
\begin{align}\label{eq:tevateeq}
\frac{t_{\rm eva}}{t_{\rm eeq}}\approx 4.7\times 10^{10} \alpha^{-1} q^{2n+3} S_{\rm f}^n \,\beta^2\left(\frac{M_{\rm f}}{1\,\rm g}\right)^2 >1\,,
\end{align}
where we used that $H_{\rm eeq}t_{\rm eeq}\sim 0.5$ (see App. \ref{App:detailedformulas}). For the case $\beta<\beta_*$, one should add an additional $q^2$ factor in the right hand side of Eq.~\eqref{eq:tevateeq}. We remind the reader that the standard case is recovered by taking the limit $q \to 1$ ($M_{\rm mb}\to M_{\rm f}$) and $n \to 0$ ($\alpha \to 1$).
It is useful to express Eq.~\eqref{eq:tevateeq} in terms of the scale factor. We find that
\begin{align}\label{eq:aevaaeeq}
\frac{a_{\rm eva}}{a_{\rm eeq}}\approx \frac{H_{\rm eeq}^{2/3}}{{2}^{1/3}
   H_{\rm eva}^{2/3}}\approx\frac{1}{2}\left(\frac{3}{2}\right)^{2/3}\frac{
   t_{\rm eva}^{2/3}}{t^{2/3}_{\rm eeq}}\,.
\end{align}
Thus, we see that the number of e-folds of the PBH dominated stage is given by
\begin{align}
\Delta N_{\rm PBH-dom}\approx \ln\left(\frac{a_{\rm eva}}{a_{\rm eeq}}\right)\approx16(1+n)+\frac{4(1+n)}{3}\ln\left(\frac{M_{\rm f}}{1 \,\rm g}\right)+\frac{2}{3}\ln\left(\frac{\beta^2 q^{3+2n}}{\alpha}\right)~\,.
\end{align}
The general conclusion is that PBH domination lasts for an additional $N\sim \tfrac{2n}{3}\ln S_{\rm f}$ number of e-folds compared to the standard PBH reheating scenario. From Eq.~\eqref{eq:tevateeq}, we may also derive the condition for PBH domination to occur (i.e.\@ ${t_{\rm eva}}>{t_{\rm eeq}}$). This yields a lower bound on $\beta$ given by
\begin{align}\label{eq:lowerboundbeta}
\beta>4.6\times 10^{-6} e^{-12n}\sqrt{\frac{\alpha}{q}}\left(\frac{M_{\rm mb}}{1\,\rm g}\right)^{-(1+n)}.
\end{align}

\begin{figure}
\centering
\includegraphics[width=\textwidth]{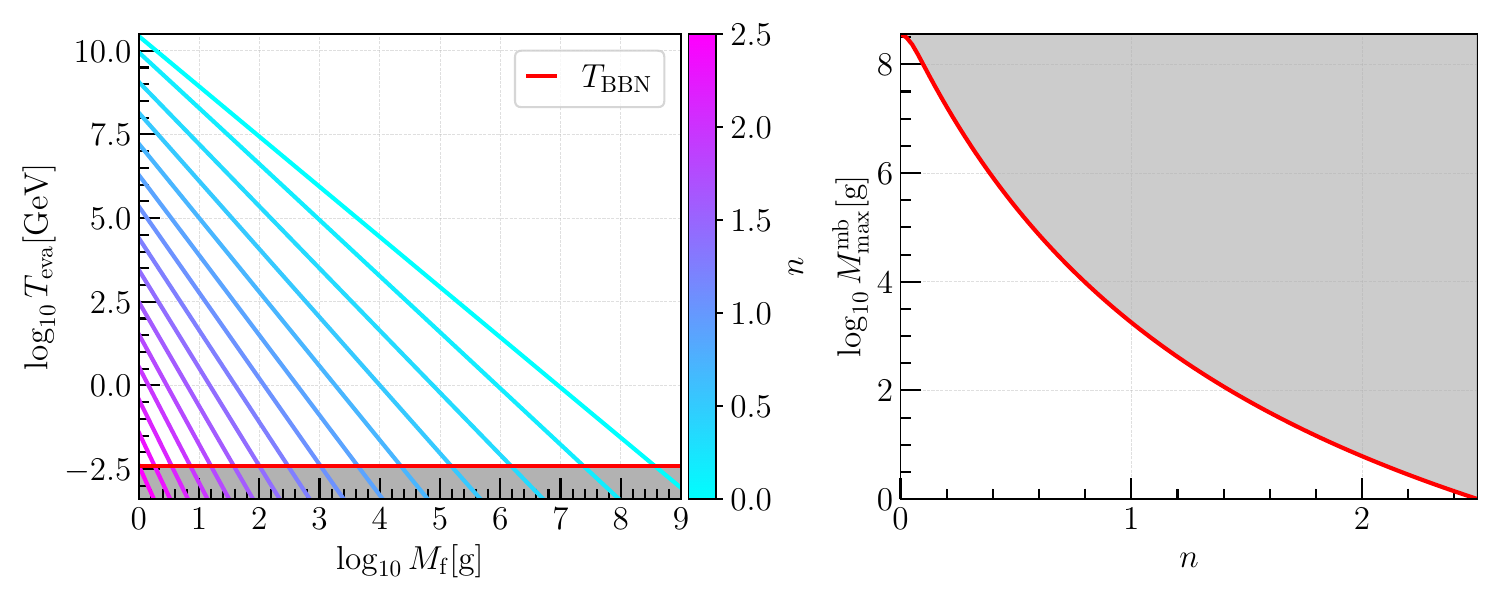}
\caption{
{\bf Left panel:} we plot the temperature at evaporation, given by Eq.~\eqref{eq:Tevamb}, as a function of $M_{\rm f}$. With the color code, we show the dependence on $n$. The red line shows the temperature at BBN, which provides the lower limit of the allowed parameter range for the temperature. The excluded parameter space is shown with a gray shaded region. We note how increasing $n$ exponentially shrinks the parameter range where PBH domination can be attained, as the evaporation is not sufficiently fast. 
{\bf Right panel:} we show the maximal mass $M^{\rm mb}_{\rm f,max}$, given by Eq.~\eqref{eq:Mfmax}, as a function of $n$. Note that we only considered the mb1 case as the results are qualitatively the same as in the mb2 case and differ only by a simple numerical factor. As in the left panel, the shaded region is excluded by BBN.}
\label{fig:Teva}
\end{figure}

As we will show, a longer PBH domination phase can enhance the GW signal significantly.
However, the duration is bounded by the requirement that the PBHs evaporate well before Big Bang nucleosynthesis (BBN) in order for the products of their evaporation to not interfere with BBN processes.
This puts a lower bound on the temperature at evaporation $T_{\rm eva} > T_{\rm BBN} \approx  4\,{\rm MeV}$ \cite{Kawasaki:1999na,Kawasaki:2000en,Hannestad:2004px,Hasegawa:2019jsa}.
Assuming PBHs dominated, we can compute the temperature of the universe at evaporation as
\begin{align}\label{eq:Tevamb}
T_{\rm eva}\approx2.75\times
   10^{10}\,{\rm GeV}\,\alpha^{1/2}e^{-12 n}\left(\frac{M_{\rm mb}}{1\,\rm g}\right)^{-(3/2+n)}\left(\frac{g_*(T_{\rm eva})}{106.75}\right)^{-1/4}~.
\end{align}
The bound on the temperature can be translated to an upper bound for the mass of the PBHs, which is given by
\begin{align}\label{eq:Mfmax}
M_{\rm f}<M^{\rm mb}_{\rm f,max}= 5 \times 10^8 \,{\rm g}\left(\frac{e^{12 n}
   q^{n+\frac{3}{2}} 7^{\frac{2n}{3}} 10^{8n}}{\alpha}\right)^{-\frac{2}{2
   n+3}} \approx 5 \times 10^8 \,{\rm g}\left(\frac{e^{31.7 n}
   q^{n+\frac{3}{2}}}{\alpha}\right)^{-\frac{2}{2
   n+3}} ~.
\end{align}
Thus, the modifications place stronger limits on the maximum mass of PBHs.
We also show this behavior in Fig.~\ref{fig:Teva}.

Lastly, we can express the BBN bound in terms of an upper bound on $n$ by enforcing the Hubble parameter at formation to be smaller than the aforementioned bound $H_{\rm f} < 2.5\times 10^{-5}\,M_{\rm pl}$ \cite{Planck:2018jri}, which provides a lower bound for the mass which is roughly given by $M_{\rm f,max}^{\rm mb}\gtrsim 1\,{\rm g}$. 
Thus, we find
\begin{align}\label{eq:nmax}
    n < \frac{59 - 3 \log q}{2(12+\log q)}
\end{align}
for mb1 (i.e.\@ $\alpha=1$).
Note that had we considered the case mb2, which includes a dependence of $\alpha$ on $n$, 
we would have found a similar bound. 
For $q=1/2$, we roughly find that $n< 2.7$. The numerical results are shown in Fig.~\ref{fig:Teva} and indicate $n< 2.5$.
If signatures of early PBH domination were discovered, this would already allow us to set bounds on the modified Hawking evaporation. We will now turn to discuss GW signatures of this scenario.

\section{Memory burden and the induced GW signal from PBH density fluctuations \label{sec:inducedGWs}}

We proceed to estimate the GW spectrum induced by PBH number density fluctuations. We will closely follow Refs.~\cite{Domenech:2021wkk,Domenech:2021ztg}, which builds upon the work of Refs.~\cite{Inomata:2020lmk,Inomata:2019ivs,Inomata:2019zqy,Papanikolaou:2020qtd,Domenech:2020ssp}, adapting the relevant calculations to the current case. The main effect of the modification to Hawking evaporation given by Eqs.~\eqref{eq:MPBHmb} is an extension of the PBH dominated era, proportional to the PBH entropy. Thus, one naively expects that the amplitude of the GW spectrum is enhanced and that the peak moves to lower frequencies with respect to the standard case.

\begin{figure}
\centering
\includegraphics[width=0.6\textwidth]{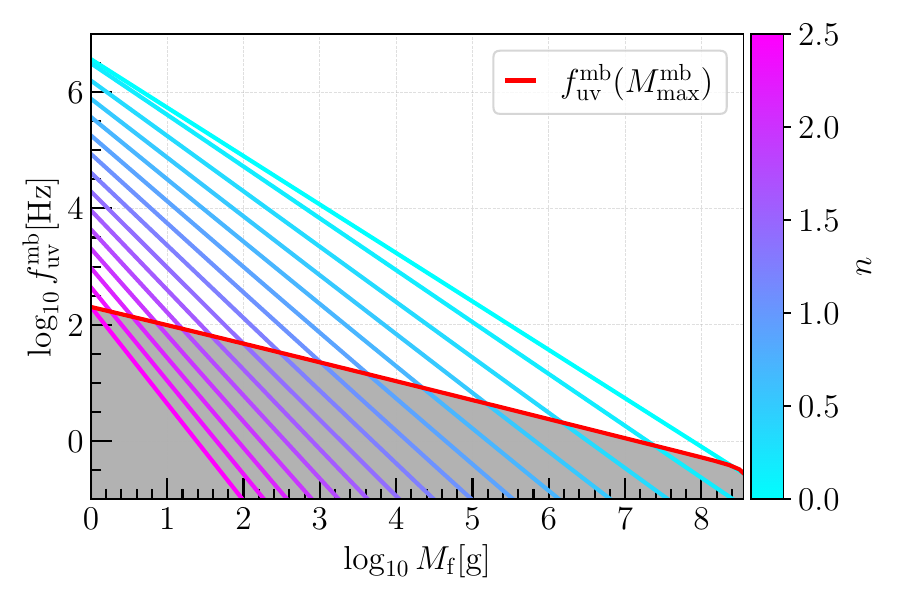}
\caption{
Peak frequency $f_{\rm uv}$ of the induced GW spectrum, given by Eq.~\eqref{eq:fuvmb}, as a function of $M_{\rm f}$. With the color code, we show how $f_{\rm uv}(M_{\rm f})$ depends on $n$. See how increasing the value of $n$ moves the peak to lower frequencies. The red line shows the minimum frequency allowed by BBN constraints, which, for a given $n$, gives a maximum mass $M_{\rm f}^{\rm max}$ given by Eq.~\eqref{eq:Mfmax}. 
While memory burden can shift the peak to lower frequencies,
it can never place it below $\sim \rm Hz$. We show the excluded parameter space with a gray shaded region.}
\label{fig:fuv}
\end{figure}

Let us first estimate the modification to the peak frequency of the induced GW spectrum. As explained in Ref.~\cite{Papanikolaou:2020qtd} (see also Refs.~\cite{Domenech:2024cjn,Domenech:2024kmh} for reviews with illustrations), the gas of PBHs has number density fluctuations that follow Poisson statistics. The peak of the density fluctuations lies at the mean distance between PBHs, where the PBH gas approximation breaks down. This is given by \cite{Papanikolaou:2020qtd}
\begin{align}\label{eq:kuv}
k_{\rm uv}=\left(\frac{4\pi n_{{\rm PBH},\rm f}a_{\rm f}}{3}\right)^{1/3}=\beta^{1/3}\gamma^{-1/3}\,k_{\rm f}\,,
\end{align}
where we used that $a_{\rm f}H_{\rm f}=k_{\rm f}$. The notation “uv” refers to the ultraviolet (or high momenta) cut-off. Interestingly, the ratio $k_{\rm uv}/k_{\rm eva}$ is independent of $\beta$ in the standard case \cite{Domenech:2020ssp}. This is because $k_{\rm eva}\propto a_{\rm eva}/a_{\rm f}\propto n_{\rm PBH,f}^{1/3}\propto \beta^{1/3}$. A similar result applies to the memory burden case since the shift in $k_{\rm eva}$ is proportional to powers of $S_{\rm f}\propto M_{\rm f}^2$, but not $\beta$. We find that
\begin{align}\label{eq:kuvkevamb}
\frac{k_{\rm uv}}{k_{\rm eva}}\approx 4989\, \alpha^{-1/3} e^{8n} \left(  \frac{M_{\rm mb}}{1\,\rm g} \right)^{\frac{2}{3}(1+n)}  \,. 
\end{align}
To find the frequency associated to $k_{\rm uv}$ today, that is $f_{\rm uv}=k_{\rm uv}/(2\pi a_0)$, we compute $f_{\rm eva}$ in terms of $T_{\rm eva}$ (by using that $k_{\rm eva}=a_{\rm eva}H_{\rm eva}$ and that $a_{\rm eva}/a_0\propto T_0/T_{\rm eva}$) and express it in terms of $M_{\rm f}$. This leads us to
\begin{align}\label{eq:fevamb}
{f_{\rm eva}}\approx 733\,{\rm Hz}\,\alpha^{1/2}e^{-12 n}\left(\frac{M_{\rm mb}}{1\,\rm g}\right)^{-3/2-n}\left(\frac{g_*(T_{\rm eva})}{106.75}\right)^{1/4}\left(\frac{g_{*s}(T_{\rm eva})}{106.75}\right)^{-1/3} \,,
\end{align}
where $g_*$ denotes the effective number of relativistic degrees of freedom and $g_{*s}$ are the effective entropic degrees of freedom.
Then, we use Eq.~\eqref{eq:kuvkevamb} to arrive at
\begin{align}\label{eq:fuvmb}
 {f_{\rm uv}}\approx  3.7\times 10^6\,{\rm Hz}\,\alpha^{1/6} e^{-4 n} \left(\frac{M_{\rm mb}}{1\,\rm g}\right)^{-\frac{5}{6}-\frac{n}{3}}\left(\frac{g_*(T_{\rm eva})}{106.75}\right)^{1/4}\left(\frac{g_{*s}(T_{\rm eva})}{106.75}\right)^{-1/3}  \,.
\end{align}
Once again, we note that we recover the standard-case expressions of Ref.~\cite{Domenech:2020ssp} respectively for $f_{\rm eva}$ and $f_{\rm uv}$ by taking $n\to 0$ ($\alpha\to 1$) and $q\to 1$ ($M_{\rm mb}\to M_{\rm f}$) in Eqs.~\eqref{eq:fevamb} and \eqref{eq:fuvmb}. From Eq.~\eqref{eq:fuvmb}, we see that the peak moves to lower frequencies mainly by a factor $e^{-4n}$ and with a different $M_{\rm f}$ scaling. However, when we take into account the BBN bound on the evaporation temperature, which sets a maximum mass for a given value of $n$ given by Eq.~\eqref{eq:Mfmax}, we find that the peak frequency $f_{\rm uv}$ is constrained to be $f_{\rm uv}\geq 0.3\,{\rm Hz}$. The equality corresponds to the standard case. We also show this result more clearly in Fig.~\ref{fig:fuv}.

We proceed to estimate the largest contribution to the induced GW spectrum. The largest production of GWs occurs right after complete evaporation when the large PBH number density fluctuations become large radiation fluctuations and create huge acoustic waves. We can directly use the analytical results of Refs.~\cite{Domenech:2021wkk,Domenech:2021ztg} by computing the modifications to the spectrum of curvature fluctuations at evaporation (which are the main source of induced GWs). We note that the analysis of Refs.~\cite{Domenech:2021wkk,Domenech:2021ztg} is still applicable as the transition from a PBH to a radiation dominated universe still occurs in much less than an e-fold. This is clear from Fig.~\ref{fig:OmegaPBHRadPlot}. While there is a longer PBH domination, the final burst of PBHs yields an almost instantaneous transition \cite{Inomata:2020lmk}. With this in mind, we define the spectrum of curvature fluctuations in the Newton gauge, which we call $\Phi$, at evaporation as
\begin{align}\label{eq:pphidust}
{\cal P}_{\Phi}(k)\Big|_{\rm eva}={\cal A}_\Phi\left(\frac{k}{k_{\rm uv}}\right)^{m}\Theta(k_{\rm uv}-k)\,,
\end{align}
which has a cut-off at $k=k_{\rm uv}$. 
We find that the amplitude of the power spectrum at evaporation is given by
\begin{align}\label{eq:Aphi}
{\cal A}_\Phi=\frac{3}{8\pi}\left(\frac{k_{\rm eeq}}{k_{\rm uv}}\right)^4{\cal S}^2_{\Phi,\rm eva}(k_{\rm uv}){\cal S}^2_{\Phi,\rm mb}(k_{\rm uv})=\frac{3q^2}{8\pi}\left(\frac{k_{\rm eeq}}{k_{\rm uv}}\right)^4\left(\frac{2}{\sqrt{3}}  \frac{\alpha}{\sqrt{{3 \alpha -1}}} \frac{k_{\rm uv}}{k_{\rm eva}} \right)^{- \frac{2}{3\alpha}}\,,
\end{align}
and the spectral slope reads
\begin{align}\label{eq:m}
     m=3-4-\frac{2}{3\alpha}=-\frac{5}{3}+\frac{2}{3}\frac{\alpha-1}{\alpha}\,.
\end{align}
Note that $k_{\rm eeq}=a_{\rm eeq}H_{\rm eeq}$ in Eq.~\eqref{eq:Aphi} is due to the decay of the curvature fluctuation before the eeq. Its ratio with $k_{\rm uv}$ reads 
\begin{align}\label{eq:kUVkeeqSC}
\frac{k_{\rm uv}}{k_{\rm eeq}}\approx2^{-1/2} \beta^{-2/3}\gamma^{-1/3} \,.
\end{align}
In the case when $\beta<\beta_*$ \eqref{eq:lower_bound_beta_standard}, one should replace $k_{\rm eeq}\to q k_{\rm eeq}$. That is, since eeq in terms of scale factor happens later by a factor $q^{-1}$, the wavenumber $k_{\rm eeq}$ is smaller by a factor $q$. We provide more details of the transfer functions from PBH formation until PBH evaporation in App.~\ref{App:evolution}. 

\begin{figure}
  \centering
  \includegraphics[width=\textwidth]{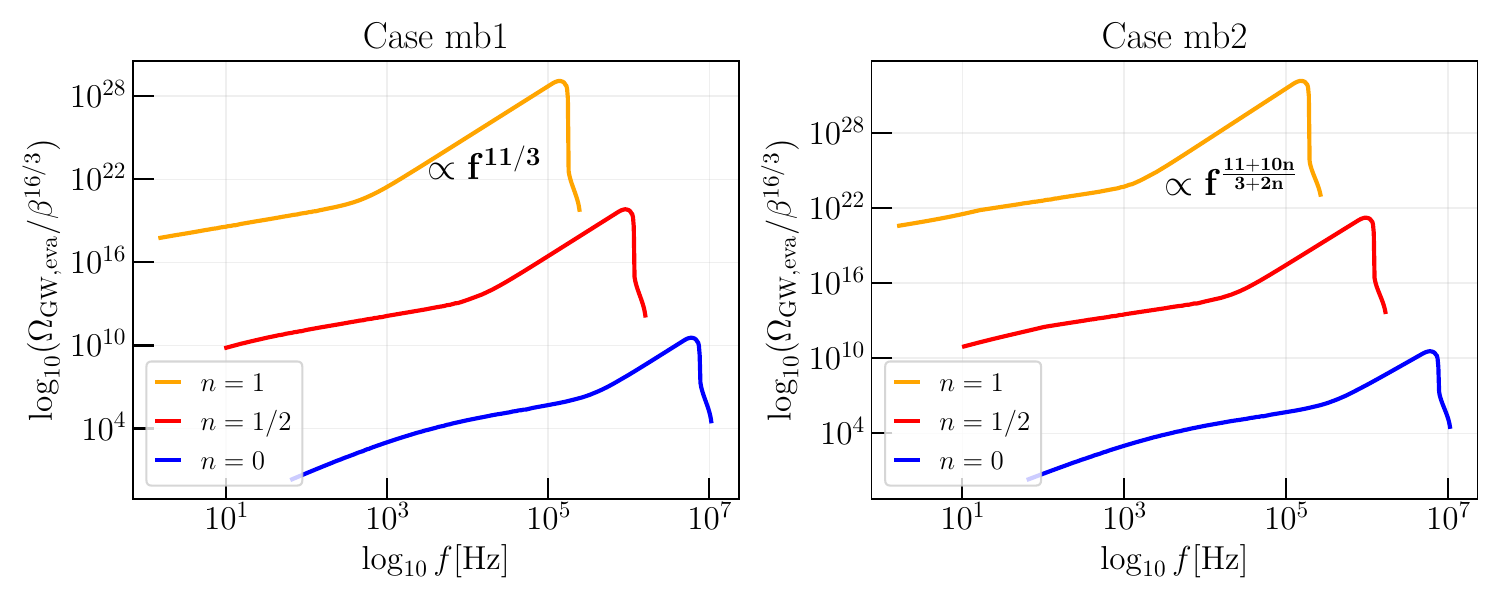}
  \caption{Comparison of the induced GW spectral densities $\Omega_{\rm GW}(f)$ for the mb1 ({\bf left panel}) and mb2 ({\bf right panel}) scenarios for an initial PBH mass of $M_{\rm  f}=1 \si{\gram}$. The spectrum peaks around $f \approx f_{\rm uv}$ and drops sharply above the resonant scale $f_{\rm res}= 2 c_s f_{\rm uv}$ before going to zero at the UV cutoff $f=2 f_{\rm uv}$. The slope of the resonant peak scales as $f^{11/3}$ or $f^{\frac{11+10n}{3+2n}}$ respectively. 
  For simplicity, we rescale the amplitude by the corresponding prefactor dependent on $\beta$, while we observe peak amplitude \eqref{eq:GWpeaknumbers} being exponentially enhanced with $n$.}
  \label{fig:OmegaGWPlot_MB1_MB2}
\end{figure}

With the power-law ansatz \eqref{eq:pphidust}, we can estimate the resonant peak in the induced GW spectrum as \cite{Domenech:2021wkk,Domenech:2021ztg}
\begin{align}\label{eq:GWspectrum}
    \Omega_{\rm GW,eva}^{\rm res}(f)\approx \Omega^{\rm peak}_{\rm GW,res}\left(\frac{f}{f_{\rm uv}}\right)^{7+2m}\int_{-s_0(f)}^{s_0(f)} ds {(1-s^2)^2}{(1-c_s^2s^2)^{m}}\,,
\end{align}
where we defined
\begin{align}\label{eq:gwspeak}
\Omega_{\rm GW,\rm res}^{\rm peak}
&\approx \frac{\pi}{3\times 2^{11}c_s}\left(1-c_s^2\right)^2\left({4c^2_s}\right)^{-m}\left(\frac{k_{\rm uv}}{k_{\rm eva}}\right)^7{\cal A}^2_{\Phi}\nonumber\\&
\approx \frac{q^4}{24576 \sqrt{3} \pi }\left(\frac{\alpha}{\sqrt{3 \alpha -1}}\right)^{-\frac{4}{3\alpha}}\left(\frac{k_{\rm eeq}}{k_{\rm uv}}\right)^8\left(\frac{k_{\rm uv}}{k_{\rm eva}}\right)^{7-\frac{4}{3\alpha}}\,,
\end{align}
and
\begin{align}\label{eq:s0}
s_0(f)=\left\{
\begin{aligned}
&1\qquad &\tfrac{f_{\rm uv}}{f}\geq\tfrac{1+c_s^{-1}}{2}\\
&2\tfrac{f_{\rm uv}}{f}-c_s^{-1}\qquad & \tfrac{1+c_s^{-1}}{2}\geq\tfrac{f_{\rm uv}}{f}\geq\tfrac{c_s^{-1}}{2}\\
&0\qquad &\tfrac{c_s^{-1}}{2}\geq\tfrac{f_{\rm uv}}{f}
\end{aligned}
\right.\,,
\end{align}
where $c_s=1/\sqrt{3}$ is the sound speed during radiation domination.
More details on the computation of the GW spectrum are given in App.~\ref{App:iGWspectrum}. We show the induced GW spectrum in Fig.~\ref{fig:OmegaGWPlot_MB1_MB2} and compare the estimation of the resonant peak with the full, numerical spectrum in Fig.~\ref{fig:OmegaGWPlotNumRes}.

\begin{figure}
\centering
\includegraphics[width=0.6\textwidth]{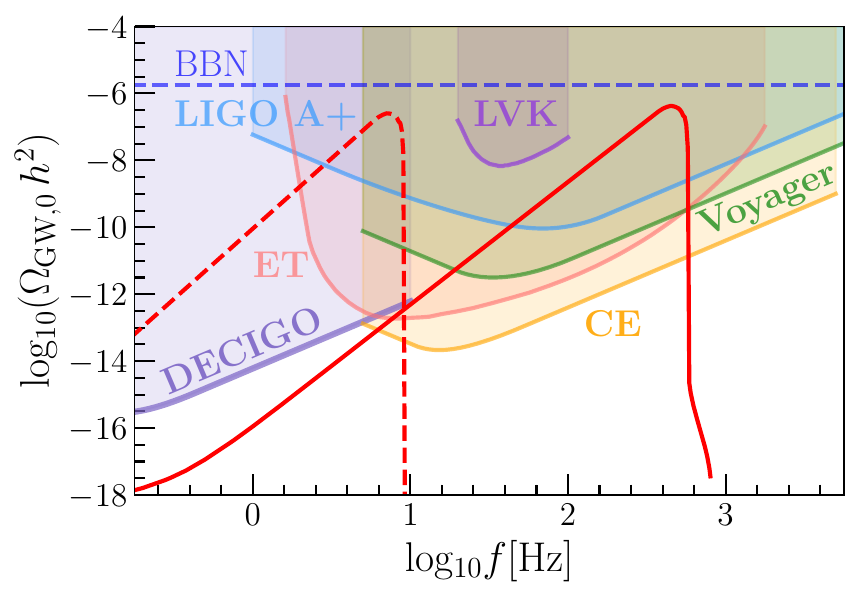}
\caption{The induced gravitational wave spectrum \eqref{eq:spectraldensitytoday} for PBH mass ${M_{\rm f}=2\times10^3\si{\gram}}$, $\beta= 10^{-8}$ and $n=\frac{1}{2}$ in the scenario mb1 ($\alpha=1$)  (solid red line), as well as the spectrum for scenario mb2 ($\alpha=1+2n/3$) with a PBH mass ${M_{\rm  f}= 6 \times 10^3\si{\gram}}$, $\beta= 5 \times 10^{-13}$ and $n=1$ (dashed red line). As one can see, for this choice of parameters, the signals enter the observational windows for several of the next-generation GW detectors. We also show the power-law integrated sensitivity curves as in Ref.~\cite{Thrane:2013oya} (but also see Ref.~\cite{Schmitz:2020syl}) for LIGO A+, Voyager, Einstein Telescope (ET) (triangular configuration with 15km arms \cite{Branchesi:2023mws}), Cosmic Explorer (CE) and DECIGO experiments (for the sensitivity curves see Refs.~\cite{ce,A+,voyager,Schmitz:2020syl,Yagi:2011wg,Kawamura:2020pcg}). In purple, we show the upper bounds from the search for an isotropic GW background from the LVK collaboration \cite{KAGRA:2021kbb}. The current (integrated) constraint from BBN \cite{Cyburt:2004yc,Arbey:2021ysg,Grohs:2023voo} is shown with a dashed blue line.
}
\label{fig:plotsensitivy}
\end{figure}

We will use $\Omega_{\rm GW,\rm res}^{\rm peak}$ from Eq.~\eqref{eq:gwspeak} as a good approximation for the peak amplitude of the GW spectrum.
It should be noted that in the case when $\beta<\beta_*$, there is an additional suppression factor $q^{8}$ coming from $k_{\rm eeq}$ in Eq.~\eqref{eq:gwspeak}. Taking $q\sim 1/2$ results in a suppression of $0.004$, which would be exactly compensated by taking a factor $3$ larger value for $\beta$.
The GW spectrum evaluated today is then given by
\begin{align}\label{eq:spectraldensitytoday}
\Omega_{\rm GW,0}h^2&\approx 1.62\times 10^{-5}\left(\frac{\Omega_{r,0}h^2}{4.18\times 10^{-5}}\right)\left(\frac{g_*(T_{\rm eva})}{106.75}\right)\left(\frac{g_{*s}(T_{\rm eva})}{106.75}\right)^{-4/3}\Omega_{\rm GW,eva}(f)\,.
\end{align}
We plot the resulting GW spectrum for some example parameters together with the sensitivity curves of some current and planned GW detectors in Fig.~\ref{fig:plotsensitivy}.

At this point, let us discuss some properties of the induced GW spectrum \eqref{eq:GWspectrum}. The most interesting modification with respect to the standard case is the possibility of changing the slope. Looking at Eq.~\eqref{eq:m}, we see that this occurs for $\alpha\neq 1$ and, therefore, this is a characteristic signature of the mb2 case (see Eq.~\eqref{eq:dMPBHscdt}). We find that for the mb2 case, the spectral index of the resonant peak is given by
\begin{align}\label{eq:gwslopemb2}
\frac{d\ln\Omega_{\rm GW}(f)}{d\ln f}\Bigg|_{\rm mb2}=\frac{11+10n}{3+2n}\,.
\end{align}
For the mb1 case, the spectral index is as in the standard case, namely $11/3$. The fact that the slope \eqref{eq:gwslopemb2} in the case mb2 is steeper is due to a faster final evaporation for $\alpha>1$ (see Eq.~\eqref{eq:MPBHmb}). From Eq.~\eqref{eq:nmax} we find that $11/3<\tfrac{d\ln\Omega_{\rm GW}(f)}{d\ln f}\lesssim 9/2$.
This tilt becomes milder in the low-frequency tail; see App.~\ref{App:evolution} for more details. 
If the GW signature of such a scenario were discovered, a measurement of the steep tilt of the spectrum would yield important information on the PBH evaporation process.

We also see from Eqs.~\eqref{eq:gwspeak} that since the ratio $k_{\rm uv}/k_{\rm eva}$ \eqref{eq:kuvkevamb} is larger, the peak amplitude of the GW spectrum is larger than in the standard case. Inserting some numbers, we find that
\begin{align}\label{eq:GWpeaknumbers}
\Omega_{\rm GW,\rm res}^{\rm peak}
\approx 1.2\times 10^{15} \,\beta ^{16/3}q^4 \left(\frac{\sqrt{2}\alpha}{\sqrt{3 \alpha -1}}\right)^{-\frac{4}{3\alpha}}\frac{
e^{\frac{8}{\alpha}\left[{ (7 \alpha -\frac{4}{3})
   n+\frac{17}{12}(\alpha-1)}\right]}}{\alpha^{\frac{7}{3}-\frac{4}{9 \alpha }}}
   \left(\frac{M_{\rm mb}}{1\,\rm g}\right)^{\frac{2 (21 \alpha -4)
   (1+n)}{9 \alpha }}\,.
\end{align}
The result of Ref.~\cite{Domenech:2020ssp} in the standard case is again recovered for $n\to 0$ ($\alpha\to 1$) and $q\to 1$ ($M_{\rm mb}\to M_{\rm f}$).
From Eq.~\eqref{eq:GWpeaknumbers}, we see that the peak amplitude \eqref{eq:GWpeaknumbers} is exponentially enhanced with $n$ and also has a stronger dependence on $M_{\rm f}$ as $n$ increases, compared with the standard case.

It is also important to understand possible degeneracy between parameters in the various cases.
Inverting the peak frequency \eqref{eq:fuvmb}, we obtain a relation $M_{\rm f}(n,f_{\rm uv})$, assuming a fixed $q$ and a fixed model (either mb1 or mb2 which can be inferred from the GW spectral slope). Thus, from $f_{\rm uv}$ we derive $M_{\rm f}(n)$. In the mb2 case, one can measure the value of $n$ from the spectral slope \eqref{eq:gwslopemb2}. Therefore, in the mb2 case, one could extract the initial fraction of PBHs from the peak amplitude \eqref{eq:GWpeaknumbers}, as one deduced $n$ and $M_{\rm f}$ from $f_{\rm uv}$ and the spectral slope. However, in the mb1 case, the parameter $n$ is not fixed by the spectral slope, and, therefore, the peak amplitude \eqref{eq:GWpeaknumbers} only provides us with a relation $\beta(n)$. In other words, there is a degeneracy between $\beta$ and $n$ in the mb1 case.

Most interestingly, despite the degeneracy in the mb1 case, we can still obtain general bounds on the initial fraction of PBHs by imposing the BBN bound on the effective number of relativistic species \cite{Cyburt:2004yc,Arbey:2021ysg} (see Ref.~\cite{Grohs:2023voo} for a recent review), which roughly states that the high redshift GW abundance should be $0.39\,\Omega_{\rm GW,eva}<0.05$ \cite{Caprini:2018mtu}. Using Eq.~\eqref{eq:GWpeaknumbers}, we conclude that
\begin{align}\label{eq:upperboundbeta}
    \beta < 1.2 \times  10^{-4}\, q^{-3/4} \frac{\alpha ^{\frac{1}{6 \alpha }+\frac{7}{16}}}{(3
   \alpha
   -1)^{\frac{1}{8\alpha} }}
   e^{\frac{-84 \alpha  n+16 n+17}{8 \alpha
   }} 
   \left(\frac{M_{\rm mb}}{1\,\rm g}\right)^{-\frac{(21 \alpha -4)
   (1+n)}{24 \alpha }} \,.
\end{align}

\begin{figure}
\centering
\includegraphics[width=\textwidth]{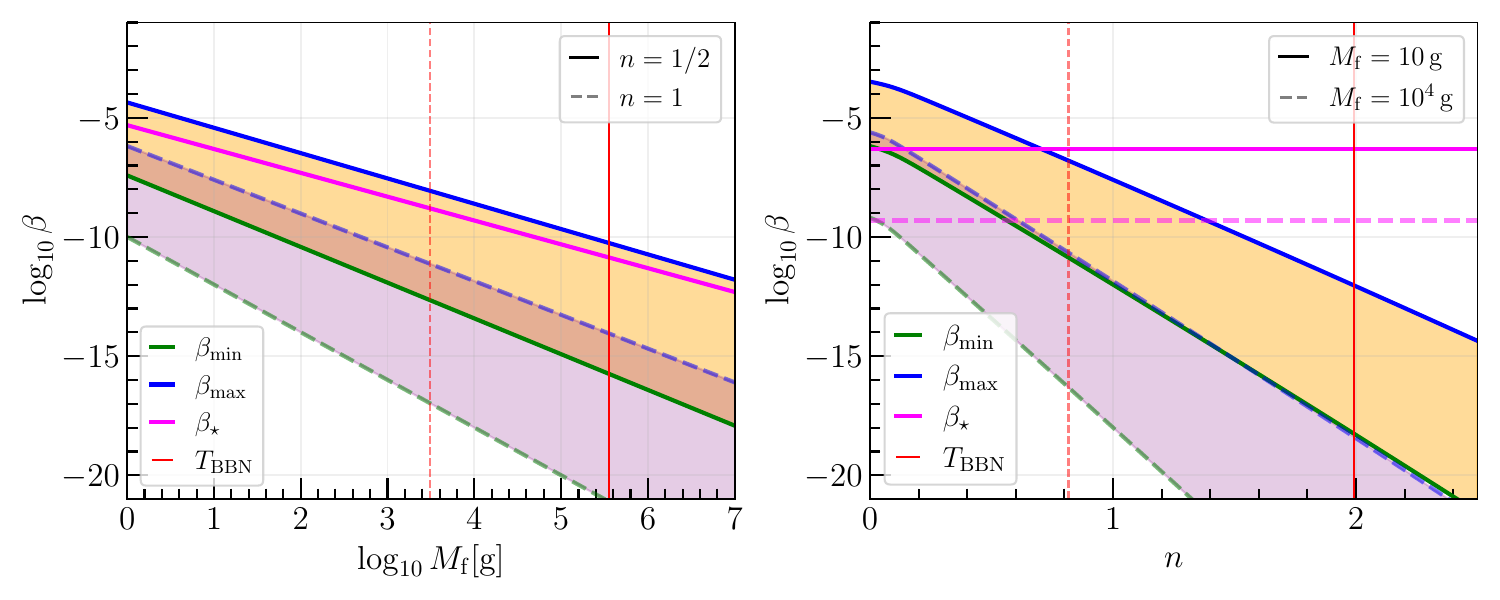}
\caption{We show the allowed parameter range for $\beta$ in the shaded regions. With blue lines, we show the upper bound from the BBN constraint on the effective number of relativistic species \eqref{eq:upperboundbeta}. With green lines, we show the lower bound \eqref{eq:lowerboundbeta} from requiring PBH domination. And with magenta lines, we show the limiting value of $\beta$ \eqref{eq:lower_bound_beta_standard} above (below) which the memory burden time occurs after (before) the early PBH-radiation equality. 
With red lines, we show the maximum value by requiring that PBH evaporate well enough before BBN, given by Eq.~\eqref{eq:Mfmax}. Thus, the parameter range to the right of the red line is excluded.
While we only show the mb1 case here, the results are qualitatively similar in the mb2 case.
{\bf Left panel:} 
we show $\beta$ as a function of $M_{\rm f}$ for two values of $n$, namely $n=\{1/2,2\}$ with solid and dashed lines, respectively. 
{\bf Right panel:}  we plot $\beta$ as a function of $n$ for two values of $M_{\rm f}$, namely $M_{\rm f}=\{10,10^4\}\rm g$ with solid and dashed line, respectively.
}
\label{fig:beta}
\end{figure}

 In Fig.~\ref{fig:beta}, we show the BBN upper bound on $\beta$ \eqref{eq:upperboundbeta} (in blue), together with the lower bound from requiring PBH domination \eqref{eq:lowerboundbeta} (in green). The main effect of the modified Hawking evaporation is that the initial fraction of PBHs must be smaller than in the standard case due to a longer evaporation time. Thus, we see that the larger the $n$, the smaller the $\beta$. We also show in magenta the limiting value of $\beta$ \eqref{eq:lower_bound_beta_standard}, which separates the case when the memory burden time \eqref{eq:tmb} is reached before or after eeq, corresponding to below or above the magenta lines, respectively. The magenta line roughly corresponds to the lower bound for PBH domination to happen in the standard case. We note how $\beta$ may decrease by several orders of magnitude with respect to the standard case.

\section{Discussion and conclusions \label{sec:conclusions}}

There are recent speculations that the standard Hawking evaporation is not valid after a black hole loses an ${\cal O}(1)$ factor of its mass \cite{Dvali2018,Dvali2020}. The so-called memory burden effect suppresses the evaporation rate by inverse powers of the black hole entropy \cite{Dvali2018,Dvali2020} (here we parameterized it as $S^{-n}$), allowing for a much larger lifetime than usually estimated for a given black hole mass. If so, PBHs with masses below $10^{10}\,\rm g$ could still explain the entire dark matter content of our universe \cite{Alexandre:2024nuo, Thoss:2024hsr} while still remaining elusive to possible discoveries.
In this work, we argue that any modification to Hawking evaporation would also significantly modify 
the scenario in which PBHs come to dominate an early phase of the universe and become responsible for its
reheating \cite{Carr:1976zz,Chapline:1976au,Lidsey:2001nj,Anantua:2008am,Hidalgo:2011fj}. In particular, we discuss how this would impact the GW signal associated with it \cite{Inomata:2020lmk,Papanikolaou:2020qtd,Domenech:2020ssp}. This provides a way to test the memory burden effect via the GW background above Hz frequencies.

We considered two models for the memory burden inspired by Ref.~\cite{Thoss:2024hsr}. One where the modification only involves a constant entropy factor (case mb1) and another where the entropy is a function of the evaporating PBH mass (case mb2). The relevant difference between these two cases is that in the mb2 case, the final burst of the PBHs happens faster than in the mb1 and the standard case. The two models are very similar otherwise, leading to a much longer stage of PBH domination. However, the faster final evaporation changes the slope of the GW spectrum of GWs induced by PBH number density fluctuations, giving a unique signature of the model mb2.

Our results can be summarized as follows. First, we find that the range of PBH masses that lead to successful PBH reheating is smaller the larger the $n$ (see Eq.~\eqref{eq:Mfmax} and Fig.~\ref{fig:Teva} for the maximum value of $M_{\rm f}$ allowed by BBN constraints on the evaporation temperature). Imposing a minimum mass of $M_{\rm f}\sim 1\,\rm g$ (related to high scale inflation) leads to an upper bound $n<2.5$ so that PBHs can successfully reheat the universe. The induced GW signal from PBH number density fluctuations gets enhanced by the longer PBH dominated stage, and the peak frequency decreases for a fixed PBH mass. However, the BBN bounds on the evaporation temperature constrain the peak frequency to $f_{\rm uv}\gtrsim \rm Hz$ (see Eq.~\eqref{eq:fuvmb} and Fig.~\ref{fig:fuv}). This means that, in general, the PBH reheating scenario leaves a high frequency signature testable by GW detectors such as the LIGO/Virgo/KAGRA collaboration, ET, CE, Voyager, and DECIGO, among others (cf.~Fig.~\ref{fig:plotsensitivy}). For the mb1 case, the spectral slope is the same as in the standard case, namely $11/3$. Because of this, it does not seem feasible to tell the mb1 case from the standard case apart, as there is a perfect degeneracy between a given value of $\beta$ and a value of $n$. One can only find a relation between the mass and $n$ using information on the peak frequency. Notably, for the mb2 case instead, the spectral slope is modified to $(11+10n)/(3+2n)$, allowing us to single out the value of $n$. For this reason, the model mb2 can be distinguished from both the mb1 and the standard case, with all the model parameters being determined by the detection of the GW background.

It is worth noting that our results summarized above apply to more general modifications of Hawking evaporation. In particular, since the modifications of Hawking evaporation are a local effect (i.e.\@, related to the individual black holes), it will not affect the $\beta$ dependence we obtain in our calculations. For instance, the peak frequency $f_{\rm uv}$ \eqref{eq:fuvmb} only depends on $M_{\rm f}$ and the peak amplitude \eqref{eq:GWpeaknumbers} will be proportional to $\beta^{16/3}$. We expect that $f_{\rm uv}$ will be bound to frequencies $\gtrsim \rm Hz$ due to BBN bounds on the evaporation temperature. Moreover, if the modification affects the time dependence of the evaporating mass (i.e.\@, not only extending it by a constant factor), the slope of the GW spectrum will be sensitive to such modification. Lastly, it is interesting to note that other effects, such as PBH spin and additional degrees of freedom (that is, increasing the value of $g_H$), only accelerate evaporation slightly. For instance, the evaporation time of highly spinning PBHs is about a factor $1/2$ smaller than that in the zero spin case \cite{Dong:2015yjs,Arbey:2019jmj}. The impact on the induced GW spectrum is thus not significant \cite{Domenech:2021wkk,Bhaumik:2022zdd}.

As extensions of this work, it would be interesting to study the impact of the memory burden on the PBH remnants as dark matter \cite{MacGibbon:1987my} (see Refs.~\cite{Chen:2014jwq,Hossenfelder:2012jw,Vidotto:2018hww,Eichhorn:2022bgu,Platania:2023srt} for details on black hole remnants) and their unique GW signature \cite{Domenech:2023mqk}, which should require smaller PBH masses for larger values of $n$. It would also be interesting to study how the spectrum of induced GWs from primordial adiabatic fluctuations, such as the ones measured by CMB observations, is enhanced due to the PBH-dominated phase as in Ref.~\cite{Inomata:2020lmk} (see also Refs.\cite{Bhaumik:2022pil,Bhaumik:2022zdd}). Our expectation is that it should be similar to the PBH number density fluctuations studied in this paper, although it might depend on whether one employs a cut-off for scales that become non-linear at evaporation \cite{Inomata:2020lmk}. It would also be interesting to explore the effect of an extended mass function in more detail \cite{Inomata:2020lmk,Papanikolaou:2022chm} and study the effect of non-Gaussianities and PBH clustering in conjuction with the memory burden effect \cite{Papanikolaou:2024kjb, He:2024luf}.
Another interesting aspect to consider is the fact that black hole thermodynamics depends on the underlying theory of gravity itself. In particular, the definition of the black hole entropy \eqref{eq:entropy} changes in theories of modified and quantum gravity \cite{Briscese:2007cd, Mureika:2015sda, Calmet:2021lny}. The induced GWs from the PBH reheating scenario could provide a unique opportunity to probe such effects, which are otherwise hard to access.
Lastly, it would be interesting to explore the modifications to the GWs from Hawking evaporation \cite{Dolgov:2011cq,Dong:2015yjs,Hooper:2019gtx,Masina:2020xhk,Masina:2021zpu,Arbey:2021ysg,Gehrman:2023esa,Ireland:2023avg}. It should also be noted that one known issue of the GWs induced by PBH density fluctuations from Refs.~\cite{Papanikolaou:2020qtd,Domenech:2020ssp}, is that some scales enter the non-linear regime before evaporation. However, curvature perturbations (which is the main source of induced GWs) remain within the regime of validity of perturbation theory, and so the calculations of the induced GW spectrum can be understood as an (optimistic) estimate. We checked in App.~\ref{App:evolution} that including the memory burden effect worsens this issue. Nevertheless, the main point of this paper is to estimate the impact of the memory burden effect on the PBH reheating scenario, which should not be affected by non-linear effects. To derive more accurate estimates, one requires more sophisticated numerical simulations.

\let\oldaddcontentsline\addcontentsline
\renewcommand{\addcontentsline}[3]{}
\begin{acknowledgments}
   We would like to thank I.~Masina and V.~Takhistov for useful feedback.
   We would also like to thank R. Haque for pointing out a typo in a previous version of the manuscript.
   SB is supported by the STFC under grant ST/X000753/1. GD, AG, and JT are supported by the DFG under the Emmy-Noether program, project number 496592360. GD is also supported by the JSPS KAKENHI grant No. JP24K00624.
\end{acknowledgments}

\let\addcontentsline\oldaddcontentsline

\appendix
\section{Detailed formulas}\label{App:detailedformulas}

In this appendix, we present the explicit expressions for the limiting masses that have been simplified in the main text by considering the limit $S_{\rm f}\gg 1$, $n\neq 0$ and $q\neq 1$. First, in Eqs.~\eqref{eq:MPBHmb} and \eqref{eq:tevamb} the parameters $M_{\rm mb}$ and $t_{\rm eva}$ should be replaced by 
\begin{align}
M_{\rm mb1}=M_{\rm f}
    \left[\frac{q^3 \left(q^{2 n}
   S_{\rm f}^n-1\right)+1}{q^{2 n}
   S_{\rm f}^{n}}\right]^{1/3}
\quad {,} \quad
t^{\rm mb1}_{\rm eva}=\left[q^3 \left(q^{2
   n} S_{\rm f}^n-1\right)+1\right]t_{\rm eva}^{\rm sc} 
\end{align}
and
\begin{align}
M_{\rm mb2}=M_{\rm f}
   \left[\frac{q^{2 n+3} S_{\rm f}^n+(1+2
   n/3)
   \left(1-q^3\right)}{
   S_{\rm f}^{{n}}}\right]^{\frac{
   1}{2 n+3}}
\quad {,} \quad
t^{\rm mb2}_{\rm eva}=\left[q^3
   \left(\frac{ q^{2 n}
   S_{\rm f}^n}{1+2n/3}-1\right)+1\right]t_{\rm eva}^{\rm sc} \,,
\end{align}
respectively for the cases mb1 and mb2.

Similar coefficients appear in the ratio $t_{\rm eva}/t_{\rm eeq}$ \eqref{eq:tevateeq}. Their more precise expression reads
\begin{align}\label{eq:tevateeqmb1}
\frac{t^{\rm mb1}_{\rm eva}}{t_{\rm eeq}}\approx 
\left [ q^3 \left(q^{2 k}
   S_{\rm f}^k-1\right)+1\right]
   \frac{t^{\rm sc}_{\rm eva}}{t^{\rm sc}_{\rm eeq}}>1\,,
\end{align}
and
\begin{align}\label{eq:tevateeqmb2}
\frac{t^{\rm mb2}_{\rm eva}}{t_{\rm eeq}}\approx
\left[q^3 \left(\frac{ q^{2 k}
   S_{\rm f}^k}{1+2 k/3}-1\right)+1\right]
   \frac{t^{\rm sc}_{\rm eva}}{t^{\rm sc}_{\rm eeq}}>1\,.
\end{align}
When replacing $t_{\rm eeq}$ by $t_{\rm eeq}^{\rm mb}$,  one obtains an additional $q^2$ factor in the right hand side of Eqs.~\eqref{eq:tevateeqmb1} and \eqref{eq:tevateeqmb2}.

Lastly, whenever possible, we use the exact expression for the scale factor in the dust-radiation dominated universe. This is given by
\begin{align}
\frac{a}{a_{\rm eeq}}=2\xi+\xi^2\,,
\end{align}
where $\xi=(\sqrt{2}-1)\tau/\tau_{\rm eeq}$. In this parametrization and  assuming $\rho_{\rm f}\approx \rho_{\rm rad,f}$, or equivalently $\beta \ll 1$, we have that
\begin{align}
{H_{\rm eeq}}\approx \sqrt{2}H_{\rm f}\frac{a_{\rm f}^2}{a_{\rm eeq}^2}\,,
\quad{\rm and}\quad 
\frac{a_{\rm f}}{a_{\rm eeq}}\approx \beta\,.
\end{align}
We can also check that $t_{\rm eeq}H_{\rm eeq}\approx \frac{4}{3}(\sqrt{2}-1)\approx 0.55$. In the main text, we approximate such relation as $t_{\rm eeq}H_{\rm eeq}\approx 1/2$ for simplicity.

\section{Evolution of curvature fluctuations induced by PBH number density fluctuations}\label{App:evolution}

Here, we give the transfer functions of the curvature fluctuation until the complete PBH evaporation is complete. The initial PBH number density fluctuations (after PBH formation) have the dimensionless spectrum given by \cite{Papanikolaou:2020qtd}
\begin{align}\label{eq:PS}
{\cal P}_{\delta_{\rm PBH,f}}(k)=\frac{2}{3\pi}\left(\frac{k}{k_{\rm uv}}\right)^3\Theta(k_{\rm uv}-k)\,.
\end{align}
The $k^3$ scaling is typical of Poisson-like fluctuations. The cut-off $k_{\rm uv}$ comes from the fact that for $k>k_{\rm uv}$ the PBH gas approximation breaks down. Such PBH number density fluctuations are isocurvature in nature. 

Now, one must follow the curvature fluctuations $\Phi$ during the radiation domination (when PBH formed) until the PBH dominated stage, where $\Phi$ becomes constant on all scales. It is known that in the PBH (matter) dominated stage, the curvature fluctuations are approximately given by
\begin{align}\label{eq:phisuperisomatter}
\Phi_{\rm iso}(a\gg a_{\rm eeq})\approx S_i\times\left\{
\begin{aligned}
&\frac{1}{5} &(k\ll k_{\rm eeq})\\
&\frac{3}{4}\left(\frac{k}{k_{\rm eeq}}\right)^{-2} &(k\gg k_{\rm eeq})
\end{aligned}
\right.\,,
\end{align}
where $S_i$ is the amplitude of the initial isocurvature fluctuations. Eq.~\eqref{eq:phisuperisomatter} provides us with the transfer function for the radiation to PBH domination transition.

We are left to compute the suppression factor due to the fact that PBH evaporation is not instantaneous. We estimate the suppression factors following the prescription given in Ref.~\cite{Inomata:2020lmk}, which uses that whenever $k\gg \Gamma$ we have that
\begin{align}\label{eq:Phisuppress}
    \frac{\Phi(t)}{\Phi_{\rm instant}}\approx \exp\left[-\int_{t_{\rm f}}^{t} d\tilde t\, \Gamma(\tilde t) \right]\,,
\end{align}
where $\Phi_{\rm instant}$ is the value under the instantaneous transition approximation. Eq.~\eqref{eq:Phisuppress} tells us that as long as the frequency of $\Phi$ is larger than the decay rate $\Gamma$, $\Phi$ tracks PBH density fluctuations and so $\Phi\propto M(t)$. $\Phi$ decouples from PBH fluctuations once $k\ll \Gamma$. Ref.~\cite{Inomata:2020lmk} provides a more accurate evaluation of the decoupling time as the time when $|\ddot\Phi|_{\rm dec}\sim \frac{k^2}{3a^2}|\Phi|_{\rm dec}$. In the set-up under study, there are two instances where curvature fluctuations experience suppression. The first time is before the memory burden time, where the mass of the PBHs goes down to $M_{\rm mb}=qM_{\rm f}$. The second time corresponds to the complete evaporation of the PBHs. Defining the suppression factor as ${\cal S}_\Phi(k)\equiv {\Phi(k)}/{\Phi_{\rm instant}(k)}$ we find, on the one hand, that due to the decay before $t_{\rm mb}$ fluctuations become suppressed by
\begin{align}
    {\cal S}_{\Phi,\rm mb}(k\gg k_{\rm mb}) \approx q\,.
\end{align}
This suppression is constant because the modes of interest, that is, $k\gg k_{\rm mb}$, never decouple, and the integral \eqref{eq:Phisuppress} gives a constant factor equal to the PBH mass lost.
On the other hand, the suppression due to the last stages of evaporation is given by
\begin{align}\label{eq:SuppressionFactorEva}
    {\cal S}_{\Phi,\rm eva}(k\gg k_{\rm eva}) \approx \left( \frac{2}{\sqrt{3}} \frac{\alpha}{\sqrt{{3 \alpha -1}}} \frac{k}{k_{\rm eva}} \right)^{- \frac{1}{3\alpha}}\,.
\end{align}
In this case, the suppression is $k$-dependent because we have that $\Gamma\to\infty$ as $t\to t_{\rm eva}$ and eventually all $k$'s decouple.
Let us note here that, in the case of a monochromatic PBH mass function, the instantaneous transition approximation with the suppression factor \eqref{eq:SuppressionFactorEva} has been shown to be in good agreement with numerical results \cite{Inomata:2020lmk}, especially for the very small scales we are most interested in. In the case of a broader mass function the transition would be more gradual, and the amplitude of the induced GWs would be slightly suppressed. The effects of an extended mass function and a gradual transition have been discussed in \cite{Papanikolaou:2022chm, Inomata:2019zqy, Pearce:2023kxp}, respectively.

In passing, we estimate the magnitude of PBH density fluctuations at evaporation. Since density fluctuations grow during the PBH-dominated era, some scales will enter the non-linear regime. To estimate this, we start with the Poisson equation, which yields
\begin{align}
\frac{\delta\rho_{\rm PBH}}{\rho_{\rm PBH}}\approx\frac{2}{3}\frac{k^2}{{\cal H}^2}\Phi\,.
\end{align}
The largest density ratio will be reached by the smallest scale fluctuation, which corresponds to $k=k_{\rm uv}$ and is given by
\begin{align}
\frac{\delta\rho_{\rm PBH}}{\rho_{\rm PBH}}(k_{\rm uv})\Bigg|_{\rm eva}\approx\frac{2}{3}\frac{k_{\rm uv}^2}{k_{\rm eva}^2}{\cal A}^{1/2}_\Phi\,.
\end{align}
We now insert the values we derived in the main text \eqref{eq:kuvkevamb}, arriving at
\begin{align}
\frac{\delta\rho_{\rm PBH}}{\rho_{\rm PBH}}(k_{\rm uv})\Bigg|_{\rm eva}\approx \sqrt{\frac{2}{3}}\left(\frac{3}{4}\right)^{\frac{1}{6 \alpha }}
   \, q \,\gamma^{2/3}\beta ^{4/3}\left(\frac{\alpha}{\sqrt{3 \alpha -1}}\right)^{-\frac{1}{3\alpha}}\frac{
    e^{\frac{(6 \alpha -1)(17+16n)}{6\alpha}}}{\alpha^{\frac{2}{3}-\frac{1}{9 \alpha }}}
   \left(\frac{M_{\rm mb}}{1\,\rm g}\right)^{\frac{2 (6\alpha-1)
   (1+n)}{9 \alpha }} \,.
\end{align}
To have an idea of the maximum value the density ratio can attain, we insert the maximum value of $\beta$ allowed by BBN constraints, Eq.~\eqref{eq:upperboundbeta}. This leads us to
\begin{align}
    \frac{\delta\rho_{\rm PBH}}{\rho_{\rm PBH}}(k_{\rm uv})\bigg|^{\rm max}_{\rm eva}\approx 25 \left(\frac{3}{4}\right)^{\frac{1}{6 \alpha }}\frac{e^{2n}}{\alpha^{1/12}} \left(\frac{M_{\rm mb}}{1\,\rm g}\right)^{\frac{1+n}{6 }} \,,
\end{align}
where for $n=1$ reads
\begin{align}
    \frac{\delta\rho_{\rm PBH}}{\rho_{\rm PBH}}(k_{\rm uv},n=1)\bigg|^{\rm max}_{\rm eva}\approx 160 \, \left(\frac{M_{\rm mb}}{1\,\rm g}\right)^{\frac{1}{3 }}\,.
\end{align}
Note that we could have been more conservative by not including the suppression factor due to Hawking evaporation, as perhaps density fluctuations might have grown more before evaporation became significant. However, calculations of the induced GWs depend on fluctuations after evaporation. For this reason, we include the suppression term. We would also like to stress that the curvature fluctuations $\Phi$, which are the source of induced GWs, always remain much smaller than unity. Thus, the estimation of the GW spectrum using the linear perturbation results is more justified, although it might overestimate the amplitude.

\section{Induced Gravitational Wave Integrals}\label{App:iGWspectrum}
In order to compute the GWs induced in the modified PBH reheating scenario, we closely follow Ref.~\cite{Domenech:2020ssp} and refer the interested reader to \cite{Inomata:2020lmk, Papanikolaou:2020qtd, Domenech:2020ssp} for more details on the computation. In particular, we will focus on the GWs induced right after PBH evaporation, which constitute the dominant contribution due to the enhancement by the fast transition to the late radiation domination (lRD) \cite{Inomata:2019ivs}.
The GW spectral density is given by
\begin{align}
    \Omega_{\rm GW} = \frac{k^2}{12 \mathcal{H}^2} \overline{\mathcal{P}_{h}}&(k,\tau) \,,
\end{align}
evaluated at a conformal time $\tau$ during lRD, when the GWs propagate as a free wave.
Here, $\mathcal{H}$ is the conformal Hubble parameter, and the overline denotes an oscillation average.
The tensor power spectrum $\overline{\mathcal{P}_{h}}$ can be calculated by
\begin{align}
    \overline{\mathcal{P}_{h}}(k,\tau)=8 \int_{0}^{\infty} dv \int_{|1-v|}^{1+v} du \left(\frac{\left(1+v^2-u^2\right)^2-4 v^2}{4 u v}\right)^2
    {\cal P}_{\delta_{\rm PBH,f}}(k u)
    {\cal P}_{\delta_{\rm PBH,f}}(k v)
    \overline{I^2}(x,u,v) \,,
    \label{eq:TensorPowerSpectrum}
\end{align}
with $x=k\tau$ and ${\cal P}_{\delta_{\rm PBH,f}}$ given in \eqref{eq:PS}. 
The dominant contribution to the kernel $\overline{I^2}$ stems from the fast oscillation of the curvature fluctuations after evaporation,
Therefore, we focus on the time derivative squared term of the full kernel. For our scales of interest, namely ${k\gg k_{\rm eeq}}$, it is given by
\begin{align}
    I_{\rm lRD}(x,u,v) \approx&
    \frac{c_s^2 u v }{32 \bar{x}} x_{\rm eva}^4
    {\cal S}_{\Phi,\rm mb}^2 {\cal S}_{\Phi,\rm eva}(u k) {\cal S}_{\Phi,\rm eva}(v k)
    T_{\Phi_{\rm iso}}(u k) T_{\Phi_{\rm iso}}(v k)
    \nonumber \\
    &\times \int_0^\infty \frac{d\tilde{x}}{\tilde{x}+x_{\rm eva}/2}
    \sin (x_1 - \tilde{x})
    \sin \left(c_s u \tilde{x}\right)
    \sin \left(c_s v \tilde{x}\right) \,,
    \label{eq:KernellRD}
\end{align}
where we defined $\bar{x}=x-x_{\rm eva}/2$, $x_1 = \bar{x}-x_{\rm eva}/2$ and $x_{\rm eva}=k\tau_{\rm eva}$.
The transfer function $T_{\Phi_{\rm iso}}$ for the curvature fluctuation is defined from \eqref{eq:phisuperisomatter} as $\Phi_{\rm iso}(k) = S_i\times T_{\Phi_{\rm iso}}(k)$, the 
suppression factor ${\cal S}_{\Phi,\rm mb}$ is given in \eqref{eq:Phisuppress} and ${\cal S}_{\Phi,\rm eva}$ is defined in \eqref{eq:SuppressionFactorEva}.
The factor ${\cal S}_{\Phi,\rm eva}(k)$ carries the modification of the scale dependence introduced by the memory burden effect.
The suppression factors and the transfer function can be summarized into ${\cal P}_{\Phi}(k)\Big|_{\rm eva}$ as in \eqref{eq:pphidust} for a more compact notation.

\begin{figure}
  \centering
  \includegraphics[width=0.6\textwidth]{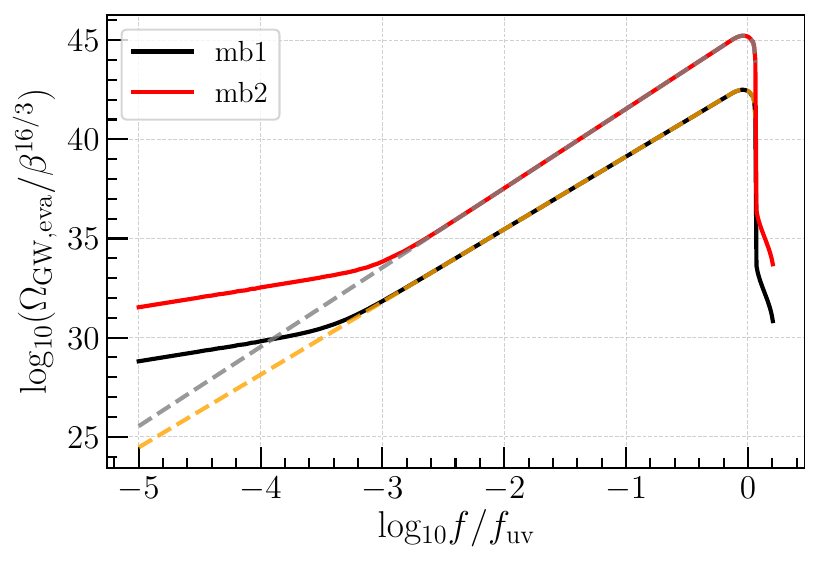}
  \caption{We compare the numerical GW spectral densities $\Omega_{\rm GW}(f)$ (solid lines), normalized by $\beta^{16/3}$, with parameters $M_{\rm PBH, f}=10^4 \si{\gram}$ and $n=\frac{1}{2}$ for both scenarios, mb1 and mb2, with the respective analytical result for the resonant contribution Eq.~\eqref{eq:GWspectrum} (dashed lines). As can be seen, the analytical approximation captures the peak very well, with deviations due to the IR tail setting in at around $f/f_{\rm uv}\sim 10^{-3}$ (for this case of $n=\frac{1}{2}$), and above the resonant scale $f_{\rm res}= 2 c_s f_{\rm uv}$ in the UV. We found that the amplitude of the numerical solution is larger than the analytical approximation \eqref{eq:GWspectrum} for the resonant peak by a factor of 2, stemming from the contribution of the non-divergent terms resulting from \eqref{eq:KernellRD}. We corrected the factor in the plot to show the good agreement of the spectral shape.}
\label{fig:OmegaGWPlotNumRes}
\end{figure}

The power spectrum $\overline{\mathcal{P}_{h}}$ has a peak near the resonant scale $u+v=c_s^{-1}$. Selecting only the dominant part of the kernel \eqref{eq:KernellRD} near this resonance yields
\begin{align}
    \overline{\mathcal{P}_{h,\rm lRD, res}}&(k,\tau,x\gg1) \approx
    \frac{ 3^{\frac{2}{3 \alpha }+2} c_s^4 q^4}{2^{\frac{4}{3 \alpha }+14}\pi ^2  \bar{x}^2}
    \left(\frac{3 \alpha -1}{\alpha ^{2}}\right)^{\frac{2}{3 \alpha }}
    \left(\frac{k}{k_{\rm uv}}\right)^6
    \left(\frac{k_{\rm eeq}}{k_{\rm eva}}\right)^8
    \left(\frac{k}{k_{\rm eva}}\right)^{-\frac{4}{3 \alpha }} \nonumber \\
    & \times \int_{k_{\rm eeq}/k}^{k_{\rm uv}/k} dv \int_{\text{max} (|1-v|,k_{\rm eeq}/k)}^{\text{min}(1+v,k_{\rm uv}/k)} du
    \frac{\left(\left(1+v^2-u^2\right)^2-4 v^2\right)^2}{(u v)^{\frac{2}{3 \alpha }+1}} \text{Ci}^2\left(| (1-(u+v)c_s)| \frac{k}{k_{\rm eva}}\right) \, ,
    \label{eq:TensorPowerSpectrumRes}
\end{align}
where the dependence on the modification due to the memory burden effect in terms of $\alpha$ becomes apparent.
Eq.~\eqref{eq:TensorPowerSpectrumRes} can be evaluated approximately near the resonant peak at $u+v=c_s^{-1}$, where the cosine integral diverges, resulting in Eq.~\eqref{eq:GWspectrum} quoted in the main text.

In order to confirm the validity of this approximation and for the spectra shown in the plots, we also solved the full momentum integrals in \eqref{eq:TensorPowerSpectrum} numerically. In this case, the integrand has the same dependence on $\alpha$ as in \eqref{eq:TensorPowerSpectrumRes}, but more cosine and sine integral terms are present, resulting from the kernel $\overline{I^2_{\rm lRD}}$ obtained from \eqref{eq:KernellRD}.
To perform the numerical integration, we first replace $k_{\rm eva}$ and $k_{\rm eeq}$ by $k_{\rm uv}$ using \eqref{eq:kuvkevamb} and \eqref{eq:kUVkeeqSC}. The double integral can then be evaluated numerically for a range of values $k/k_{\rm uv}$ by first performing the $u$-integral on a grid of $v$-values, interpolating the result to obtain the integrand as a function of $v$, and then performing the $v$-integral.
Examples of the resulting induced GW spectra are shown in Fig.~\ref{fig:OmegaGWPlotNumRes}.

\let\oldaddcontentsline\addcontentsline
\renewcommand{\addcontentsline}[3]{}
\bibliography{main}

\begin{thebibliography}{97}%
\makeatletter
\providecommand \@ifxundefined [1]{%
 \@ifx{#1\undefined}
}%
\providecommand \@ifnum [1]{%
 \ifnum #1\expandafter \@firstoftwo
 \else \expandafter \@secondoftwo
 \fi
}%
\providecommand \@ifx [1]{%
 \ifx #1\expandafter \@firstoftwo
 \else \expandafter \@secondoftwo
 \fi
}%
\providecommand \natexlab [1]{#1}%
\providecommand \enquote  [1]{``#1''}%
\providecommand \bibnamefont  [1]{#1}%
\providecommand \bibfnamefont [1]{#1}%
\providecommand \citenamefont [1]{#1}%
\providecommand \href@noop [0]{\@secondoftwo}%
\providecommand \href [0]{\begingroup \@sanitize@url \@href}%
\providecommand \@href[1]{\@@startlink{#1}\@@href}%
\providecommand \@@href[1]{\endgroup#1\@@endlink}%
\providecommand \@sanitize@url [0]{\catcode `\\12\catcode `\$12\catcode
  `\&12\catcode `\#12\catcode `\^12\catcode `\_12\catcode `\%12\relax}%
\providecommand \@@startlink[1]{}%
\providecommand \@@endlink[0]{}%
\providecommand \url  [0]{\begingroup\@sanitize@url \@url }%
\providecommand \@url [1]{\endgroup\@href {#1}{\urlprefix }}%
\providecommand \urlprefix  [0]{URL }%
\providecommand \Eprint [0]{\href }%
\providecommand \doibase [0]{https://doi.org/}%
\providecommand \selectlanguage [0]{\@gobble}%
\providecommand \bibinfo  [0]{\@secondoftwo}%
\providecommand \bibfield  [0]{\@secondoftwo}%
\providecommand \translation [1]{[#1]}%
\providecommand \BibitemOpen [0]{}%
\providecommand \bibitemStop [0]{}%
\providecommand \bibitemNoStop [0]{.\EOS\space}%
\providecommand \EOS [0]{\spacefactor3000\relax}%
\providecommand \BibitemShut  [1]{\csname bibitem#1\endcsname}%
\let\auto@bib@innerbib\@empty
\bibitem [{\citenamefont {Zel'dovich}\ and\ \citenamefont
  {Novikov}(1967)}]{Zeldovich1967}%
  \BibitemOpen
  \bibfield  {author} {\bibinfo {author} {\bibfnamefont {Y.~B.}\ \bibnamefont
  {Zel'dovich}}\ and\ \bibinfo {author} {\bibfnamefont {I.~D.}\ \bibnamefont
  {Novikov}},\ }\href@noop {} {\bibfield  {journal} {\bibinfo  {journal} {Sov.
  Astron.}\ }\textbf {\bibinfo {volume} {10}},\ \bibinfo {pages} {602}
  (\bibinfo {year} {1967})}\BibitemShut {NoStop}%
\bibitem [{\citenamefont {Hawking}(1971)}]{Hawking1971}%
  \BibitemOpen
  \bibfield  {author} {\bibinfo {author} {\bibfnamefont {S.}~\bibnamefont
  {Hawking}},\ }\href {https://doi.org/10.1093/mnras/152.1.75} {\bibfield
  {journal} {\bibinfo  {journal} {Mon. Not. Roy. Astron. Soc.}\ }\textbf
  {\bibinfo {volume} {152}},\ \bibinfo {pages} {75} (\bibinfo {year}
  {1971})}\BibitemShut {NoStop}%
\bibitem [{\citenamefont {Carr}\ and\ \citenamefont
  {Hawking}(1974)}]{Carr1974}%
  \BibitemOpen
  \bibfield  {author} {\bibinfo {author} {\bibfnamefont {B.~J.}\ \bibnamefont
  {Carr}}\ and\ \bibinfo {author} {\bibfnamefont {S.~W.}\ \bibnamefont
  {Hawking}},\ }\href {https://doi.org/10.1093/mnras/168.2.399} {\bibfield
  {journal} {\bibinfo  {journal} {Mon. Not. Roy. Astron. Soc.}\ }\textbf
  {\bibinfo {volume} {168}},\ \bibinfo {pages} {399} (\bibinfo {year}
  {1974})}\BibitemShut {NoStop}%
\bibitem [{\citenamefont {Chapline}(1975)}]{Chapline1975}%
  \BibitemOpen
  \bibfield  {author} {\bibinfo {author} {\bibfnamefont {G.~F.}\ \bibnamefont
  {Chapline}},\ }\href {https://doi.org/10.1038/253251a0} {\bibfield  {journal}
  {\bibinfo  {journal} {Nature}\ }\textbf {\bibinfo {volume} {253}},\ \bibinfo
  {pages} {251} (\bibinfo {year} {1975})}\BibitemShut {NoStop}%
\bibitem [{\citenamefont {Khlopov}(2010)}]{Khlopov:2008qy}%
  \BibitemOpen
  \bibfield  {author} {\bibinfo {author} {\bibfnamefont {M.~Y.}\ \bibnamefont
  {Khlopov}},\ }\href {https://doi.org/10.1088/1674-4527/10/6/001} {\bibfield
  {journal} {\bibinfo  {journal} {Res. Astron. Astrophys.}\ }\textbf {\bibinfo
  {volume} {10}},\ \bibinfo {pages} {495} (\bibinfo {year} {2010})},\ \Eprint
  {https://arxiv.org/abs/0801.0116} {arXiv:0801.0116 [astro-ph]} \BibitemShut
  {NoStop}%
\bibitem [{\citenamefont {Sasaki}\ \emph {et~al.}(2018)\citenamefont {Sasaki},
  \citenamefont {Suyama}, \citenamefont {Tanaka},\ and\ \citenamefont
  {Yokoyama}}]{Sasaki:2018dmp}%
  \BibitemOpen
  \bibfield  {author} {\bibinfo {author} {\bibfnamefont {M.}~\bibnamefont
  {Sasaki}}, \bibinfo {author} {\bibfnamefont {T.}~\bibnamefont {Suyama}},
  \bibinfo {author} {\bibfnamefont {T.}~\bibnamefont {Tanaka}},\ and\ \bibinfo
  {author} {\bibfnamefont {S.}~\bibnamefont {Yokoyama}},\ }\href
  {https://doi.org/10.1088/1361-6382/aaa7b4} {\bibfield  {journal} {\bibinfo
  {journal} {Class. Quant. Grav.}\ }\textbf {\bibinfo {volume} {35}},\ \bibinfo
  {pages} {063001} (\bibinfo {year} {2018})},\ \Eprint
  {https://arxiv.org/abs/1801.05235} {arXiv:1801.05235 [astro-ph.CO]}
  \BibitemShut {NoStop}%
\bibitem [{\citenamefont {Carr}\ \emph {et~al.}(2021)\citenamefont {Carr},
  \citenamefont {Kohri}, \citenamefont {Sendouda},\ and\ \citenamefont
  {Yokoyama}}]{Carr:2020gox}%
  \BibitemOpen
  \bibfield  {author} {\bibinfo {author} {\bibfnamefont {B.}~\bibnamefont
  {Carr}}, \bibinfo {author} {\bibfnamefont {K.}~\bibnamefont {Kohri}},
  \bibinfo {author} {\bibfnamefont {Y.}~\bibnamefont {Sendouda}},\ and\
  \bibinfo {author} {\bibfnamefont {J.}~\bibnamefont {Yokoyama}},\ }\href
  {https://doi.org/10.1088/1361-6633/ac1e31} {\bibfield  {journal} {\bibinfo
  {journal} {Rept. Prog. Phys.}\ }\textbf {\bibinfo {volume} {84}},\ \bibinfo
  {pages} {116902} (\bibinfo {year} {2021})},\ \Eprint
  {https://arxiv.org/abs/2002.12778} {arXiv:2002.12778 [astro-ph.CO]}
  \BibitemShut {NoStop}%
\bibitem [{\citenamefont {Green}\ and\ \citenamefont
  {Kavanagh}(2021)}]{Green:2020jor}%
  \BibitemOpen
  \bibfield  {author} {\bibinfo {author} {\bibfnamefont {A.~M.}\ \bibnamefont
  {Green}}\ and\ \bibinfo {author} {\bibfnamefont {B.~J.}\ \bibnamefont
  {Kavanagh}},\ }\href {https://doi.org/10.1088/1361-6471/abc534} {\bibfield
  {journal} {\bibinfo  {journal} {J. Phys. G}\ }\textbf {\bibinfo {volume}
  {48}},\ \bibinfo {pages} {4} (\bibinfo {year} {2021})},\ \Eprint
  {https://arxiv.org/abs/2007.10722} {arXiv:2007.10722 [astro-ph.CO]}
  \BibitemShut {NoStop}%
\bibitem [{\citenamefont {Escriv\`a}\ \emph {et~al.}(2022)\citenamefont
  {Escriv\`a}, \citenamefont {Kuhnel},\ and\ \citenamefont
  {Tada}}]{Escriva:2022duf}%
  \BibitemOpen
  \bibfield  {author} {\bibinfo {author} {\bibfnamefont {A.}~\bibnamefont
  {Escriv\`a}}, \bibinfo {author} {\bibfnamefont {F.}~\bibnamefont {Kuhnel}},\
  and\ \bibinfo {author} {\bibfnamefont {Y.}~\bibnamefont {Tada}},\ }\href@noop
  {} {\  (\bibinfo {year} {2022})},\ \Eprint {https://arxiv.org/abs/2211.05767}
  {arXiv:2211.05767 [astro-ph.CO]} \BibitemShut {NoStop}%
\bibitem [{\citenamefont {\"Ozsoy}\ and\ \citenamefont
  {Tasinato}(2023)}]{Ozsoy:2023ryl}%
  \BibitemOpen
  \bibfield  {author} {\bibinfo {author} {\bibfnamefont {O.}~\bibnamefont
  {\"Ozsoy}}\ and\ \bibinfo {author} {\bibfnamefont {G.}~\bibnamefont
  {Tasinato}},\ }\href {https://doi.org/10.3390/universe9050203} {\bibfield
  {journal} {\bibinfo  {journal} {Universe}\ }\textbf {\bibinfo {volume} {9}},\
  \bibinfo {pages} {203} (\bibinfo {year} {2023})},\ \Eprint
  {https://arxiv.org/abs/2301.03600} {arXiv:2301.03600 [astro-ph.CO]}
  \BibitemShut {NoStop}%
\bibitem [{\citenamefont {Hawking}(1974)}]{Hawking1974}%
  \BibitemOpen
  \bibfield  {author} {\bibinfo {author} {\bibfnamefont {S.~W.}\ \bibnamefont
  {Hawking}},\ }\href {https://doi.org/10.1038/248030a0} {\bibfield  {journal}
  {\bibinfo  {journal} {Nature}\ }\textbf {\bibinfo {volume} {248}},\ \bibinfo
  {pages} {30} (\bibinfo {year} {1974})}\BibitemShut {NoStop}%
\bibitem [{\citenamefont {de~Freitas~Pacheco}\ \emph
  {et~al.}(2023)\citenamefont {de~Freitas~Pacheco}, \citenamefont {Kiritsis},
  \citenamefont {Lucca},\ and\ \citenamefont
  {Silk}}]{deFreitasPacheco:2023hpb}%
  \BibitemOpen
  \bibfield  {author} {\bibinfo {author} {\bibfnamefont {J.~A.}\ \bibnamefont
  {de~Freitas~Pacheco}}, \bibinfo {author} {\bibfnamefont {E.}~\bibnamefont
  {Kiritsis}}, \bibinfo {author} {\bibfnamefont {M.}~\bibnamefont {Lucca}},\
  and\ \bibinfo {author} {\bibfnamefont {J.}~\bibnamefont {Silk}},\ }\href
  {https://doi.org/10.1103/PhysRevD.107.123525} {\bibfield  {journal} {\bibinfo
   {journal} {Phys. Rev. D}\ }\textbf {\bibinfo {volume} {107}},\ \bibinfo
  {pages} {123525} (\bibinfo {year} {2023})},\ \Eprint
  {https://arxiv.org/abs/2301.13215} {arXiv:2301.13215 [astro-ph.CO]}
  \BibitemShut {NoStop}%
\bibitem [{\citenamefont {Dvali}\ \emph {et~al.}(2022)\citenamefont {Dvali},
  \citenamefont {K\"uhnel},\ and\ \citenamefont {Zantedeschi}}]{Dvali:2021ofp}%
  \BibitemOpen
  \bibfield  {author} {\bibinfo {author} {\bibfnamefont {G.}~\bibnamefont
  {Dvali}}, \bibinfo {author} {\bibfnamefont {F.}~\bibnamefont {K\"uhnel}},\
  and\ \bibinfo {author} {\bibfnamefont {M.}~\bibnamefont {Zantedeschi}},\
  }\href {https://doi.org/10.1103/PhysRevLett.129.061302} {\bibfield  {journal}
  {\bibinfo  {journal} {Phys. Rev. Lett.}\ }\textbf {\bibinfo {volume} {129}},\
  \bibinfo {pages} {061302} (\bibinfo {year} {2022})},\ \Eprint
  {https://arxiv.org/abs/2112.08354} {arXiv:2112.08354 [hep-th]} \BibitemShut
  {NoStop}%
\bibitem [{\citenamefont {Friedlander}\ \emph {et~al.}(2022)\citenamefont
  {Friedlander}, \citenamefont {Mack}, \citenamefont {Schon}, \citenamefont
  {Song},\ and\ \citenamefont {Vincent}}]{Extradims}%
  \BibitemOpen
  \bibfield  {author} {\bibinfo {author} {\bibfnamefont {A.}~\bibnamefont
  {Friedlander}}, \bibinfo {author} {\bibfnamefont {K.~J.}\ \bibnamefont
  {Mack}}, \bibinfo {author} {\bibfnamefont {S.}~\bibnamefont {Schon}},
  \bibinfo {author} {\bibfnamefont {N.}~\bibnamefont {Song}},\ and\ \bibinfo
  {author} {\bibfnamefont {A.~C.}\ \bibnamefont {Vincent}},\ }\href
  {https://doi.org/10.1103/PhysRevD.105.103508} {\bibfield  {journal} {\bibinfo
   {journal} {Phys. Rev. D}\ }\textbf {\bibinfo {volume} {105}},\ \bibinfo
  {pages} {103508} (\bibinfo {year} {2022})},\ \Eprint
  {https://arxiv.org/abs/2201.11761} {arXiv:2201.11761 [hep-ph]} \BibitemShut
  {NoStop}%
\bibitem [{\citenamefont {Anchordoqui}\ \emph {et~al.}(2022)\citenamefont
  {Anchordoqui}, \citenamefont {Antoniadis},\ and\ \citenamefont
  {Lust}}]{Extradims2}%
  \BibitemOpen
  \bibfield  {author} {\bibinfo {author} {\bibfnamefont {L.~A.}\ \bibnamefont
  {Anchordoqui}}, \bibinfo {author} {\bibfnamefont {I.}~\bibnamefont
  {Antoniadis}},\ and\ \bibinfo {author} {\bibfnamefont {D.}~\bibnamefont
  {Lust}},\ }\href {https://doi.org/10.1103/PhysRevD.106.086001} {\bibfield
  {journal} {\bibinfo  {journal} {Phys. Rev. D}\ }\textbf {\bibinfo {volume}
  {106}},\ \bibinfo {pages} {086001} (\bibinfo {year} {2022})},\ \Eprint
  {https://arxiv.org/abs/2206.07071} {arXiv:2206.07071 [hep-th]} \BibitemShut
  {NoStop}%
\bibitem [{\citenamefont {Anchordoqui}\ \emph {et~al.}(2024)\citenamefont
  {Anchordoqui}, \citenamefont {Antoniadis},\ and\ \citenamefont
  {Lust}}]{Anchordoqui:2024dxu}%
  \BibitemOpen
  \bibfield  {author} {\bibinfo {author} {\bibfnamefont {L.~A.}\ \bibnamefont
  {Anchordoqui}}, \bibinfo {author} {\bibfnamefont {I.}~\bibnamefont
  {Antoniadis}},\ and\ \bibinfo {author} {\bibfnamefont {D.}~\bibnamefont
  {Lust}},\ }\href@noop {} {\  (\bibinfo {year} {2024})},\ \Eprint
  {https://arxiv.org/abs/2403.19604} {arXiv:2403.19604 [hep-th]} \BibitemShut
  {NoStop}%
\bibitem [{\citenamefont {Arkani-Hamed}\ \emph {et~al.}(1998)\citenamefont
  {Arkani-Hamed}, \citenamefont {Dimopoulos},\ and\ \citenamefont
  {Dvali}}]{ADD}%
  \BibitemOpen
  \bibfield  {author} {\bibinfo {author} {\bibfnamefont {N.}~\bibnamefont
  {Arkani-Hamed}}, \bibinfo {author} {\bibfnamefont {S.}~\bibnamefont
  {Dimopoulos}},\ and\ \bibinfo {author} {\bibfnamefont {G.~R.}\ \bibnamefont
  {Dvali}},\ }\href {https://doi.org/10.1016/S0370-2693(98)00466-3} {\bibfield
  {journal} {\bibinfo  {journal} {Phys. Lett. B}\ }\textbf {\bibinfo {volume}
  {429}},\ \bibinfo {pages} {263} (\bibinfo {year} {1998})},\ \Eprint
  {https://arxiv.org/abs/hep-ph/9803315} {arXiv:hep-ph/9803315} \BibitemShut
  {NoStop}%
\bibitem [{\citenamefont {Dvali}\ \emph {et~al.}(2020)\citenamefont {Dvali},
  \citenamefont {Eisemann}, \citenamefont {Michel},\ and\ \citenamefont
  {Zell}}]{Dvali2020}%
  \BibitemOpen
  \bibfield  {author} {\bibinfo {author} {\bibfnamefont {G.}~\bibnamefont
  {Dvali}}, \bibinfo {author} {\bibfnamefont {L.}~\bibnamefont {Eisemann}},
  \bibinfo {author} {\bibfnamefont {M.}~\bibnamefont {Michel}},\ and\ \bibinfo
  {author} {\bibfnamefont {S.}~\bibnamefont {Zell}},\ }\href
  {https://doi.org/10.1103/PhysRevD.102.103523} {\bibfield  {journal} {\bibinfo
   {journal} {Phys. Rev. D}\ }\textbf {\bibinfo {volume} {102}},\ \bibinfo
  {pages} {103523} (\bibinfo {year} {2020})},\ \Eprint
  {https://arxiv.org/abs/2006.00011} {arXiv:2006.00011 [hep-th]} \BibitemShut
  {NoStop}%
\bibitem [{\citenamefont {Dvali}(2018)}]{Dvali2018}%
  \BibitemOpen
  \bibfield  {author} {\bibinfo {author} {\bibfnamefont {G.}~\bibnamefont
  {Dvali}},\ }\href@noop {} {\  (\bibinfo {year} {2018})},\ \Eprint
  {https://arxiv.org/abs/1810.02336} {arXiv:1810.02336 [hep-th]} \BibitemShut
  {NoStop}%
\bibitem [{\citenamefont {Dvali}\ \emph {et~al.}(2021)\citenamefont {Dvali},
  \citenamefont {K\"uhnel},\ and\ \citenamefont {Zantedeschi}}]{Dvali:2021byy}%
  \BibitemOpen
  \bibfield  {author} {\bibinfo {author} {\bibfnamefont {G.}~\bibnamefont
  {Dvali}}, \bibinfo {author} {\bibfnamefont {F.}~\bibnamefont {K\"uhnel}},\
  and\ \bibinfo {author} {\bibfnamefont {M.}~\bibnamefont {Zantedeschi}},\
  }\href {https://doi.org/10.1103/PhysRevD.104.123507} {\bibfield  {journal}
  {\bibinfo  {journal} {Phys. Rev. D}\ }\textbf {\bibinfo {volume} {104}},\
  \bibinfo {pages} {123507} (\bibinfo {year} {2021})},\ \Eprint
  {https://arxiv.org/abs/2108.09471} {arXiv:2108.09471 [hep-ph]} \BibitemShut
  {NoStop}%
\bibitem [{\citenamefont {Alexandre}\ \emph {et~al.}(2024)\citenamefont
  {Alexandre}, \citenamefont {Dvali},\ and\ \citenamefont
  {Koutsangelas}}]{Alexandre:2024nuo}%
  \BibitemOpen
  \bibfield  {author} {\bibinfo {author} {\bibfnamefont {A.}~\bibnamefont
  {Alexandre}}, \bibinfo {author} {\bibfnamefont {G.}~\bibnamefont {Dvali}},\
  and\ \bibinfo {author} {\bibfnamefont {E.}~\bibnamefont {Koutsangelas}},\
  }\href@noop {} {\  (\bibinfo {year} {2024})},\ \Eprint
  {https://arxiv.org/abs/2402.14069} {arXiv:2402.14069 [hep-ph]} \BibitemShut
  {NoStop}%
\bibitem [{\citenamefont {Thoss}\ \emph {et~al.}(2024)\citenamefont {Thoss},
  \citenamefont {Burkert},\ and\ \citenamefont {Kohri}}]{Thoss:2024hsr}%
  \BibitemOpen
  \bibfield  {author} {\bibinfo {author} {\bibfnamefont {V.}~\bibnamefont
  {Thoss}}, \bibinfo {author} {\bibfnamefont {A.}~\bibnamefont {Burkert}},\
  and\ \bibinfo {author} {\bibfnamefont {K.}~\bibnamefont {Kohri}},\
  }\href@noop {} {\  (\bibinfo {year} {2024})},\ \Eprint
  {https://arxiv.org/abs/2402.17823} {arXiv:2402.17823 [astro-ph.CO]}
  \BibitemShut {NoStop}%
\bibitem [{\citenamefont {Carr}(1976)}]{Carr:1976zz}%
  \BibitemOpen
  \bibfield  {author} {\bibinfo {author} {\bibfnamefont {B.~J.}\ \bibnamefont
  {Carr}},\ }\href {https://doi.org/10.1086/154351} {\bibfield  {journal}
  {\bibinfo  {journal} {Astrophys. J.}\ }\textbf {\bibinfo {volume} {206}},\
  \bibinfo {pages} {8} (\bibinfo {year} {1976})}\BibitemShut {NoStop}%
\bibitem [{\citenamefont {Chapline}(1976)}]{Chapline:1976au}%
  \BibitemOpen
  \bibfield  {author} {\bibinfo {author} {\bibfnamefont {G.~F.}\ \bibnamefont
  {Chapline}},\ }\href {https://doi.org/10.1038/261550a0} {\bibfield  {journal}
  {\bibinfo  {journal} {Nature}\ }\textbf {\bibinfo {volume} {261}},\ \bibinfo
  {pages} {550} (\bibinfo {year} {1976})}\BibitemShut {NoStop}%
\bibitem [{\citenamefont {Garcia-Bellido}\ \emph {et~al.}(1996)\citenamefont
  {Garcia-Bellido}, \citenamefont {Linde},\ and\ \citenamefont
  {Wands}}]{Garcia-Bellido:1996mdl}%
  \BibitemOpen
  \bibfield  {author} {\bibinfo {author} {\bibfnamefont {J.}~\bibnamefont
  {Garcia-Bellido}}, \bibinfo {author} {\bibfnamefont {A.~D.}\ \bibnamefont
  {Linde}},\ and\ \bibinfo {author} {\bibfnamefont {D.}~\bibnamefont {Wands}},\
  }\href {https://doi.org/10.1103/PhysRevD.54.6040} {\bibfield  {journal}
  {\bibinfo  {journal} {Phys. Rev. D}\ }\textbf {\bibinfo {volume} {54}},\
  \bibinfo {pages} {6040} (\bibinfo {year} {1996})},\ \Eprint
  {https://arxiv.org/abs/astro-ph/9605094} {arXiv:astro-ph/9605094}
  \BibitemShut {NoStop}%
\bibitem [{\citenamefont {Lidsey}\ \emph {et~al.}(2002)\citenamefont {Lidsey},
  \citenamefont {Matos},\ and\ \citenamefont {Urena-Lopez}}]{Lidsey:2001nj}%
  \BibitemOpen
  \bibfield  {author} {\bibinfo {author} {\bibfnamefont {J.~E.}\ \bibnamefont
  {Lidsey}}, \bibinfo {author} {\bibfnamefont {T.}~\bibnamefont {Matos}},\ and\
  \bibinfo {author} {\bibfnamefont {L.~A.}\ \bibnamefont {Urena-Lopez}},\
  }\href {https://doi.org/10.1103/PhysRevD.66.023514} {\bibfield  {journal}
  {\bibinfo  {journal} {Phys. Rev. D}\ }\textbf {\bibinfo {volume} {66}},\
  \bibinfo {pages} {023514} (\bibinfo {year} {2002})},\ \Eprint
  {https://arxiv.org/abs/astro-ph/0111292} {arXiv:astro-ph/0111292}
  \BibitemShut {NoStop}%
\bibitem [{\citenamefont {Anantua}\ \emph {et~al.}(2009)\citenamefont
  {Anantua}, \citenamefont {Easther},\ and\ \citenamefont
  {Giblin}}]{Anantua:2008am}%
  \BibitemOpen
  \bibfield  {author} {\bibinfo {author} {\bibfnamefont {R.}~\bibnamefont
  {Anantua}}, \bibinfo {author} {\bibfnamefont {R.}~\bibnamefont {Easther}},\
  and\ \bibinfo {author} {\bibfnamefont {J.~T.}\ \bibnamefont {Giblin}},\
  }\href {https://doi.org/10.1103/PhysRevLett.103.111303} {\bibfield  {journal}
  {\bibinfo  {journal} {Phys. Rev. Lett.}\ }\textbf {\bibinfo {volume} {103}},\
  \bibinfo {pages} {111303} (\bibinfo {year} {2009})},\ \Eprint
  {https://arxiv.org/abs/0812.0825} {arXiv:0812.0825 [astro-ph]} \BibitemShut
  {NoStop}%
\bibitem [{\citenamefont {Hidalgo}\ \emph {et~al.}(2012)\citenamefont
  {Hidalgo}, \citenamefont {Urena-Lopez},\ and\ \citenamefont
  {Liddle}}]{Hidalgo:2011fj}%
  \BibitemOpen
  \bibfield  {author} {\bibinfo {author} {\bibfnamefont {J.~C.}\ \bibnamefont
  {Hidalgo}}, \bibinfo {author} {\bibfnamefont {L.~A.}\ \bibnamefont
  {Urena-Lopez}},\ and\ \bibinfo {author} {\bibfnamefont {A.~R.}\ \bibnamefont
  {Liddle}},\ }\href {https://doi.org/10.1103/PhysRevD.85.044055} {\bibfield
  {journal} {\bibinfo  {journal} {Phys. Rev. D}\ }\textbf {\bibinfo {volume}
  {85}},\ \bibinfo {pages} {044055} (\bibinfo {year} {2012})},\ \Eprint
  {https://arxiv.org/abs/1107.5669} {arXiv:1107.5669 [astro-ph.CO]}
  \BibitemShut {NoStop}%
\bibitem [{\citenamefont {Riajul~Haque}\ \emph {et~al.}(2023)\citenamefont
  {Riajul~Haque}, \citenamefont {Kpatcha}, \citenamefont {Maity},\ and\
  \citenamefont {Mambrini}}]{RiajulHaque:2023cqe}%
  \BibitemOpen
  \bibfield  {author} {\bibinfo {author} {\bibfnamefont {M.}~\bibnamefont
  {Riajul~Haque}}, \bibinfo {author} {\bibfnamefont {E.}~\bibnamefont
  {Kpatcha}}, \bibinfo {author} {\bibfnamefont {D.}~\bibnamefont {Maity}},\
  and\ \bibinfo {author} {\bibfnamefont {Y.}~\bibnamefont {Mambrini}},\ }\href
  {https://doi.org/10.1103/PhysRevD.108.063523} {\bibfield  {journal} {\bibinfo
   {journal} {Phys. Rev. D}\ }\textbf {\bibinfo {volume} {108}},\ \bibinfo
  {pages} {063523} (\bibinfo {year} {2023})},\ \Eprint
  {https://arxiv.org/abs/2305.10518} {arXiv:2305.10518 [hep-ph]} \BibitemShut
  {NoStop}%
\bibitem [{\citenamefont {Inomata}\ \emph {et~al.}(2020)\citenamefont
  {Inomata}, \citenamefont {Kawasaki}, \citenamefont {Mukaida}, \citenamefont
  {Terada},\ and\ \citenamefont {Yanagida}}]{Inomata:2020lmk}%
  \BibitemOpen
  \bibfield  {author} {\bibinfo {author} {\bibfnamefont {K.}~\bibnamefont
  {Inomata}}, \bibinfo {author} {\bibfnamefont {M.}~\bibnamefont {Kawasaki}},
  \bibinfo {author} {\bibfnamefont {K.}~\bibnamefont {Mukaida}}, \bibinfo
  {author} {\bibfnamefont {T.}~\bibnamefont {Terada}},\ and\ \bibinfo {author}
  {\bibfnamefont {T.~T.}\ \bibnamefont {Yanagida}},\ }\href
  {https://doi.org/10.1103/PhysRevD.101.123533} {\bibfield  {journal} {\bibinfo
   {journal} {Phys. Rev. D}\ }\textbf {\bibinfo {volume} {101}},\ \bibinfo
  {pages} {123533} (\bibinfo {year} {2020})},\ \Eprint
  {https://arxiv.org/abs/2003.10455} {arXiv:2003.10455 [astro-ph.CO]}
  \BibitemShut {NoStop}%
\bibitem [{\citenamefont {Papanikolaou}\ \emph {et~al.}(2021)\citenamefont
  {Papanikolaou}, \citenamefont {Vennin},\ and\ \citenamefont
  {Langlois}}]{Papanikolaou:2020qtd}%
  \BibitemOpen
  \bibfield  {author} {\bibinfo {author} {\bibfnamefont {T.}~\bibnamefont
  {Papanikolaou}}, \bibinfo {author} {\bibfnamefont {V.}~\bibnamefont
  {Vennin}},\ and\ \bibinfo {author} {\bibfnamefont {D.}~\bibnamefont
  {Langlois}},\ }\href {https://doi.org/10.1088/1475-7516/2021/03/053}
  {\bibfield  {journal} {\bibinfo  {journal} {JCAP}\ }\textbf {\bibinfo
  {volume} {03}},\ \bibinfo {pages} {053}},\ \Eprint
  {https://arxiv.org/abs/2010.11573} {arXiv:2010.11573 [astro-ph.CO]}
  \BibitemShut {NoStop}%
\bibitem [{\citenamefont {Dom\`enech}\ \emph
  {et~al.}(2021{\natexlab{a}})\citenamefont {Dom\`enech}, \citenamefont {Lin},\
  and\ \citenamefont {Sasaki}}]{Domenech:2020ssp}%
  \BibitemOpen
  \bibfield  {author} {\bibinfo {author} {\bibfnamefont {G.}~\bibnamefont
  {Dom\`enech}}, \bibinfo {author} {\bibfnamefont {C.}~\bibnamefont {Lin}},\
  and\ \bibinfo {author} {\bibfnamefont {M.}~\bibnamefont {Sasaki}},\ }\href
  {https://doi.org/10.1088/1475-7516/2021/11/E01} {\bibfield  {journal}
  {\bibinfo  {journal} {JCAP}\ }\textbf {\bibinfo {volume} {04}},\ \bibinfo
  {pages} {062}},\ \bibinfo {note} {[Erratum: JCAP 11, E01 (2021)]},\ \Eprint
  {https://arxiv.org/abs/2012.08151} {arXiv:2012.08151 [gr-qc]} \BibitemShut
  {NoStop}%
\bibitem [{\citenamefont {Dom\`enech}\ \emph
  {et~al.}(2021{\natexlab{b}})\citenamefont {Dom\`enech}, \citenamefont
  {Takhistov},\ and\ \citenamefont {Sasaki}}]{Domenech:2021wkk}%
  \BibitemOpen
  \bibfield  {author} {\bibinfo {author} {\bibfnamefont {G.}~\bibnamefont
  {Dom\`enech}}, \bibinfo {author} {\bibfnamefont {V.}~\bibnamefont
  {Takhistov}},\ and\ \bibinfo {author} {\bibfnamefont {M.}~\bibnamefont
  {Sasaki}},\ }\href {https://doi.org/10.1016/j.physletb.2021.136722}
  {\bibfield  {journal} {\bibinfo  {journal} {Phys. Lett. B}\ }\textbf
  {\bibinfo {volume} {823}},\ \bibinfo {pages} {136722} (\bibinfo {year}
  {2021}{\natexlab{b}})},\ \Eprint {https://arxiv.org/abs/2105.06816}
  {arXiv:2105.06816 [astro-ph.CO]} \BibitemShut {NoStop}%
\bibitem [{\citenamefont {Papanikolaou}\ \emph {et~al.}(2022)\citenamefont
  {Papanikolaou}, \citenamefont {Tzerefos}, \citenamefont {Basilakos},\ and\
  \citenamefont {Saridakis}}]{Papanikolaou:2021uhe}%
  \BibitemOpen
  \bibfield  {author} {\bibinfo {author} {\bibfnamefont {T.}~\bibnamefont
  {Papanikolaou}}, \bibinfo {author} {\bibfnamefont {C.}~\bibnamefont
  {Tzerefos}}, \bibinfo {author} {\bibfnamefont {S.}~\bibnamefont
  {Basilakos}},\ and\ \bibinfo {author} {\bibfnamefont {E.~N.}\ \bibnamefont
  {Saridakis}},\ }\href {https://doi.org/10.1088/1475-7516/2022/10/013}
  {\bibfield  {journal} {\bibinfo  {journal} {JCAP}\ }\textbf {\bibinfo
  {volume} {10}},\ \bibinfo {pages} {013}},\ \Eprint
  {https://arxiv.org/abs/2112.15059} {arXiv:2112.15059 [astro-ph.CO]}
  \BibitemShut {NoStop}%
\bibitem [{\citenamefont {Papanikolaou}(2022)}]{Papanikolaou:2022chm}%
  \BibitemOpen
  \bibfield  {author} {\bibinfo {author} {\bibfnamefont {T.}~\bibnamefont
  {Papanikolaou}},\ }\href {https://doi.org/10.1088/1475-7516/2022/10/089}
  {\bibfield  {journal} {\bibinfo  {journal} {JCAP}\ }\textbf {\bibinfo
  {volume} {10}},\ \bibinfo {pages} {089}},\ \Eprint
  {https://arxiv.org/abs/2207.11041} {arXiv:2207.11041 [astro-ph.CO]}
  \BibitemShut {NoStop}%
\bibitem [{\citenamefont {Pearce}\ \emph {et~al.}(2023)\citenamefont {Pearce},
  \citenamefont {Pearce}, \citenamefont {White},\ and\ \citenamefont
  {Bal\'azs}}]{Pearce:2023kxp}%
  \BibitemOpen
  \bibfield  {author} {\bibinfo {author} {\bibfnamefont {M.}~\bibnamefont
  {Pearce}}, \bibinfo {author} {\bibfnamefont {L.}~\bibnamefont {Pearce}},
  \bibinfo {author} {\bibfnamefont {G.}~\bibnamefont {White}},\ and\ \bibinfo
  {author} {\bibfnamefont {C.}~\bibnamefont {Bal\'azs}},\ }\href@noop {} {\
  (\bibinfo {year} {2023})},\ \Eprint {https://arxiv.org/abs/2311.12340}
  {arXiv:2311.12340 [astro-ph.CO]} \BibitemShut {NoStop}%
\bibitem [{\citenamefont {Bhaumik}\ \emph {et~al.}(2022)\citenamefont
  {Bhaumik}, \citenamefont {Ghoshal},\ and\ \citenamefont
  {Lewicki}}]{Bhaumik:2022pil}%
  \BibitemOpen
  \bibfield  {author} {\bibinfo {author} {\bibfnamefont {N.}~\bibnamefont
  {Bhaumik}}, \bibinfo {author} {\bibfnamefont {A.}~\bibnamefont {Ghoshal}},\
  and\ \bibinfo {author} {\bibfnamefont {M.}~\bibnamefont {Lewicki}},\ }\href
  {https://doi.org/10.1007/JHEP07(2022)130} {\bibfield  {journal} {\bibinfo
  {journal} {JHEP}\ }\textbf {\bibinfo {volume} {07}},\ \bibinfo {pages}
  {130}},\ \Eprint {https://arxiv.org/abs/2205.06260} {arXiv:2205.06260
  [astro-ph.CO]} \BibitemShut {NoStop}%
\bibitem [{\citenamefont {Bhaumik}\ \emph {et~al.}(2023)\citenamefont
  {Bhaumik}, \citenamefont {Ghoshal}, \citenamefont {Jain},\ and\ \citenamefont
  {Lewicki}}]{Bhaumik:2022zdd}%
  \BibitemOpen
  \bibfield  {author} {\bibinfo {author} {\bibfnamefont {N.}~\bibnamefont
  {Bhaumik}}, \bibinfo {author} {\bibfnamefont {A.}~\bibnamefont {Ghoshal}},
  \bibinfo {author} {\bibfnamefont {R.~K.}\ \bibnamefont {Jain}},\ and\
  \bibinfo {author} {\bibfnamefont {M.}~\bibnamefont {Lewicki}},\ }\href
  {https://doi.org/10.1007/JHEP05(2023)169} {\bibfield  {journal} {\bibinfo
  {journal} {JHEP}\ }\textbf {\bibinfo {volume} {05}},\ \bibinfo {pages}
  {169}},\ \Eprint {https://arxiv.org/abs/2212.00775} {arXiv:2212.00775
  [astro-ph.CO]} \BibitemShut {NoStop}%
\bibitem [{\citenamefont {Dom\`enech}(2021)}]{Domenech:2021ztg}%
  \BibitemOpen
  \bibfield  {author} {\bibinfo {author} {\bibfnamefont {G.}~\bibnamefont
  {Dom\`enech}},\ }\href {https://doi.org/10.3390/universe7110398} {\bibfield
  {journal} {\bibinfo  {journal} {Universe}\ }\textbf {\bibinfo {volume} {7}},\
  \bibinfo {pages} {398} (\bibinfo {year} {2021})},\ \Eprint
  {https://arxiv.org/abs/2109.01398} {arXiv:2109.01398 [gr-qc]} \BibitemShut
  {NoStop}%
\bibitem [{\citenamefont {Dom\`enech}(2024{\natexlab{a}})}]{Domenech:2023jve}%
  \BibitemOpen
  \bibfield  {author} {\bibinfo {author} {\bibfnamefont {G.}~\bibnamefont
  {Dom\`enech}},\ }\href {https://doi.org/10.1007/s43673-023-00109-z}
  {\bibfield  {journal} {\bibinfo  {journal} {AAPPS Bull.}\ }\textbf {\bibinfo
  {volume} {34}},\ \bibinfo {pages} {4} (\bibinfo {year}
  {2024}{\natexlab{a}})},\ \Eprint {https://arxiv.org/abs/2311.02065}
  {arXiv:2311.02065 [gr-qc]} \BibitemShut {NoStop}%
\bibitem [{\citenamefont {Franciolini}\ and\ \citenamefont
  {Pani}(2023)}]{Franciolini:2023osw}%
  \BibitemOpen
  \bibfield  {author} {\bibinfo {author} {\bibfnamefont {G.}~\bibnamefont
  {Franciolini}}\ and\ \bibinfo {author} {\bibfnamefont {P.}~\bibnamefont
  {Pani}},\ }\href {https://doi.org/10.1103/PhysRevD.108.083527} {\bibfield
  {journal} {\bibinfo  {journal} {Phys. Rev. D}\ }\textbf {\bibinfo {volume}
  {108}},\ \bibinfo {pages} {083527} (\bibinfo {year} {2023})},\ \Eprint
  {https://arxiv.org/abs/2304.13576} {arXiv:2304.13576 [astro-ph.CO]}
  \BibitemShut {NoStop}%
\bibitem [{\citenamefont {Aggarwal}\ \emph {et~al.}(2021)\citenamefont
  {Aggarwal} \emph {et~al.}}]{Aggarwal:2020olq}%
  \BibitemOpen
  \bibfield  {author} {\bibinfo {author} {\bibfnamefont {N.}~\bibnamefont
  {Aggarwal}} \emph {et~al.},\ }\href
  {https://doi.org/10.1007/s41114-021-00032-5} {\bibfield  {journal} {\bibinfo
  {journal} {Living Rev. Rel.}\ }\textbf {\bibinfo {volume} {24}},\ \bibinfo
  {pages} {4} (\bibinfo {year} {2021})},\ \Eprint
  {https://arxiv.org/abs/2011.12414} {arXiv:2011.12414 [gr-qc]} \BibitemShut
  {NoStop}%
\bibitem [{\citenamefont {Franciolini}\ \emph {et~al.}(2022)\citenamefont
  {Franciolini}, \citenamefont {Maharana},\ and\ \citenamefont
  {Muia}}]{Franciolini:2022htd}%
  \BibitemOpen
  \bibfield  {author} {\bibinfo {author} {\bibfnamefont {G.}~\bibnamefont
  {Franciolini}}, \bibinfo {author} {\bibfnamefont {A.}~\bibnamefont
  {Maharana}},\ and\ \bibinfo {author} {\bibfnamefont {F.}~\bibnamefont
  {Muia}},\ }\href {https://doi.org/10.1103/PhysRevD.106.103520} {\bibfield
  {journal} {\bibinfo  {journal} {Phys. Rev. D}\ }\textbf {\bibinfo {volume}
  {106}},\ \bibinfo {pages} {103520} (\bibinfo {year} {2022})},\ \Eprint
  {https://arxiv.org/abs/2205.02153} {arXiv:2205.02153 [astro-ph.CO]}
  \BibitemShut {NoStop}%
\bibitem [{\citenamefont {Hooper}\ \emph {et~al.}(2019)\citenamefont {Hooper},
  \citenamefont {Krnjaic},\ and\ \citenamefont {McDermott}}]{Hooper:2019gtx}%
  \BibitemOpen
  \bibfield  {author} {\bibinfo {author} {\bibfnamefont {D.}~\bibnamefont
  {Hooper}}, \bibinfo {author} {\bibfnamefont {G.}~\bibnamefont {Krnjaic}},\
  and\ \bibinfo {author} {\bibfnamefont {S.~D.}\ \bibnamefont {McDermott}},\
  }\href {https://doi.org/10.1007/JHEP08(2019)001} {\bibfield  {journal}
  {\bibinfo  {journal} {JHEP}\ }\textbf {\bibinfo {volume} {08}},\ \bibinfo
  {pages} {001}},\ \Eprint {https://arxiv.org/abs/1905.01301} {arXiv:1905.01301
  [hep-ph]} \BibitemShut {NoStop}%
\bibitem [{\citenamefont {Masina}(2020)}]{Masina:2020xhk}%
  \BibitemOpen
  \bibfield  {author} {\bibinfo {author} {\bibfnamefont {I.}~\bibnamefont
  {Masina}},\ }\href {https://doi.org/10.1140/epjp/s13360-020-00564-9}
  {\bibfield  {journal} {\bibinfo  {journal} {Eur. Phys. J. Plus}\ }\textbf
  {\bibinfo {volume} {135}},\ \bibinfo {pages} {552} (\bibinfo {year}
  {2020})},\ \Eprint {https://arxiv.org/abs/2004.04740} {arXiv:2004.04740
  [hep-ph]} \BibitemShut {NoStop}%
\bibitem [{\citenamefont {Wang}(2024)}]{Wang:2023wsm}%
  \BibitemOpen
  \bibfield  {author} {\bibinfo {author} {\bibfnamefont {S.-J.}\ \bibnamefont
  {Wang}},\ }\href {https://doi.org/10.1209/0295-5075/ad3b36} {\bibfield
  {journal} {\bibinfo  {journal} {EPL}\ }\textbf {\bibinfo {volume} {146}},\
  \bibinfo {pages} {39002} (\bibinfo {year} {2024})},\ \Eprint
  {https://arxiv.org/abs/2308.07909} {arXiv:2308.07909 [gr-qc]} \BibitemShut
  {NoStop}%
\bibitem [{\citenamefont {Crawford}\ and\ \citenamefont
  {Schramm}(1982)}]{Crawford:1982yz}%
  \BibitemOpen
  \bibfield  {author} {\bibinfo {author} {\bibfnamefont {M.}~\bibnamefont
  {Crawford}}\ and\ \bibinfo {author} {\bibfnamefont {D.~N.}\ \bibnamefont
  {Schramm}},\ }\href {https://doi.org/10.1038/298538a0} {\bibfield  {journal}
  {\bibinfo  {journal} {Nature}\ }\textbf {\bibinfo {volume} {298}},\ \bibinfo
  {pages} {538} (\bibinfo {year} {1982})}\BibitemShut {NoStop}%
\bibitem [{\citenamefont {Kodama}\ \emph {et~al.}(1982)\citenamefont {Kodama},
  \citenamefont {Sasaki},\ and\ \citenamefont {Sato}}]{Kodama:1982sf}%
  \BibitemOpen
  \bibfield  {author} {\bibinfo {author} {\bibfnamefont {H.}~\bibnamefont
  {Kodama}}, \bibinfo {author} {\bibfnamefont {M.}~\bibnamefont {Sasaki}},\
  and\ \bibinfo {author} {\bibfnamefont {K.}~\bibnamefont {Sato}},\ }\href
  {https://doi.org/10.1143/PTP.68.1979} {\bibfield  {journal} {\bibinfo
  {journal} {Prog. Theor. Phys.}\ }\textbf {\bibinfo {volume} {68}},\ \bibinfo
  {pages} {1979} (\bibinfo {year} {1982})}\BibitemShut {NoStop}%
\bibitem [{\citenamefont {Flores}\ \emph {et~al.}(2024)\citenamefont {Flores},
  \citenamefont {Kusenko},\ and\ \citenamefont {Sasaki}}]{Flores:2024lng}%
  \BibitemOpen
  \bibfield  {author} {\bibinfo {author} {\bibfnamefont {M.~M.}\ \bibnamefont
  {Flores}}, \bibinfo {author} {\bibfnamefont {A.}~\bibnamefont {Kusenko}},\
  and\ \bibinfo {author} {\bibfnamefont {M.}~\bibnamefont {Sasaki}},\
  }\href@noop {} {\  (\bibinfo {year} {2024})},\ \Eprint
  {https://arxiv.org/abs/2402.13341} {arXiv:2402.13341 [hep-ph]} \BibitemShut
  {NoStop}%
\bibitem [{\citenamefont {Cotner}\ and\ \citenamefont
  {Kusenko}(2017)}]{Cotner:2016cvr}%
  \BibitemOpen
  \bibfield  {author} {\bibinfo {author} {\bibfnamefont {E.}~\bibnamefont
  {Cotner}}\ and\ \bibinfo {author} {\bibfnamefont {A.}~\bibnamefont
  {Kusenko}},\ }\href {https://doi.org/10.1103/PhysRevLett.119.031103}
  {\bibfield  {journal} {\bibinfo  {journal} {Phys. Rev. Lett.}\ }\textbf
  {\bibinfo {volume} {119}},\ \bibinfo {pages} {031103} (\bibinfo {year}
  {2017})},\ \Eprint {https://arxiv.org/abs/1612.02529} {arXiv:1612.02529
  [astro-ph.CO]} \BibitemShut {NoStop}%
\bibitem [{\citenamefont {Cotner}\ \emph {et~al.}(2019)\citenamefont {Cotner},
  \citenamefont {Kusenko}, \citenamefont {Sasaki},\ and\ \citenamefont
  {Takhistov}}]{Cotner:2019ykd}%
  \BibitemOpen
  \bibfield  {author} {\bibinfo {author} {\bibfnamefont {E.}~\bibnamefont
  {Cotner}}, \bibinfo {author} {\bibfnamefont {A.}~\bibnamefont {Kusenko}},
  \bibinfo {author} {\bibfnamefont {M.}~\bibnamefont {Sasaki}},\ and\ \bibinfo
  {author} {\bibfnamefont {V.}~\bibnamefont {Takhistov}},\ }\href
  {https://doi.org/10.1088/1475-7516/2019/10/077} {\bibfield  {journal}
  {\bibinfo  {journal} {JCAP}\ }\textbf {\bibinfo {volume} {10}},\ \bibinfo
  {pages} {077}},\ \Eprint {https://arxiv.org/abs/1907.10613} {arXiv:1907.10613
  [astro-ph.CO]} \BibitemShut {NoStop}%
\bibitem [{\citenamefont {Flores}\ and\ \citenamefont
  {Kusenko}(2023)}]{Flores:2021jas}%
  \BibitemOpen
  \bibfield  {author} {\bibinfo {author} {\bibfnamefont {M.~M.}\ \bibnamefont
  {Flores}}\ and\ \bibinfo {author} {\bibfnamefont {A.}~\bibnamefont
  {Kusenko}},\ }\href {https://doi.org/10.1088/1475-7516/2023/05/013}
  {\bibfield  {journal} {\bibinfo  {journal} {JCAP}\ }\textbf {\bibinfo
  {volume} {05}},\ \bibinfo {pages} {013}},\ \Eprint
  {https://arxiv.org/abs/2108.08416} {arXiv:2108.08416 [hep-ph]} \BibitemShut
  {NoStop}%
\bibitem [{\citenamefont {Amendola}\ \emph {et~al.}(2018)\citenamefont
  {Amendola}, \citenamefont {Rubio},\ and\ \citenamefont
  {Wetterich}}]{Amendola:2017xhl}%
  \BibitemOpen
  \bibfield  {author} {\bibinfo {author} {\bibfnamefont {L.}~\bibnamefont
  {Amendola}}, \bibinfo {author} {\bibfnamefont {J.}~\bibnamefont {Rubio}},\
  and\ \bibinfo {author} {\bibfnamefont {C.}~\bibnamefont {Wetterich}},\ }\href
  {https://doi.org/10.1103/PhysRevD.97.081302} {\bibfield  {journal} {\bibinfo
  {journal} {Phys. Rev. D}\ }\textbf {\bibinfo {volume} {97}},\ \bibinfo
  {pages} {081302} (\bibinfo {year} {2018})},\ \Eprint
  {https://arxiv.org/abs/1711.09915} {arXiv:1711.09915 [astro-ph.CO]}
  \BibitemShut {NoStop}%
\bibitem [{\citenamefont {Flores}\ and\ \citenamefont
  {Kusenko}(2021)}]{Flores:2020drq}%
  \BibitemOpen
  \bibfield  {author} {\bibinfo {author} {\bibfnamefont {M.~M.}\ \bibnamefont
  {Flores}}\ and\ \bibinfo {author} {\bibfnamefont {A.}~\bibnamefont
  {Kusenko}},\ }\href {https://doi.org/10.1103/PhysRevLett.126.041101}
  {\bibfield  {journal} {\bibinfo  {journal} {Phys. Rev. Lett.}\ }\textbf
  {\bibinfo {volume} {126}},\ \bibinfo {pages} {041101} (\bibinfo {year}
  {2021})},\ \Eprint {https://arxiv.org/abs/2008.12456} {arXiv:2008.12456
  [astro-ph.CO]} \BibitemShut {NoStop}%
\bibitem [{\citenamefont {Dom\`enech}\ \emph {et~al.}(2023)\citenamefont
  {Dom\`enech}, \citenamefont {Inman}, \citenamefont {Kusenko},\ and\
  \citenamefont {Sasaki}}]{Domenech:2023afs}%
  \BibitemOpen
  \bibfield  {author} {\bibinfo {author} {\bibfnamefont {G.}~\bibnamefont
  {Dom\`enech}}, \bibinfo {author} {\bibfnamefont {D.}~\bibnamefont {Inman}},
  \bibinfo {author} {\bibfnamefont {A.}~\bibnamefont {Kusenko}},\ and\ \bibinfo
  {author} {\bibfnamefont {M.}~\bibnamefont {Sasaki}},\ }\href
  {https://doi.org/10.1103/PhysRevD.108.103543} {\bibfield  {journal} {\bibinfo
   {journal} {Phys. Rev. D}\ }\textbf {\bibinfo {volume} {108}},\ \bibinfo
  {pages} {103543} (\bibinfo {year} {2023})},\ \Eprint
  {https://arxiv.org/abs/2304.13053} {arXiv:2304.13053 [astro-ph.CO]}
  \BibitemShut {NoStop}%
\bibitem [{\citenamefont {Martin}\ \emph {et~al.}(2020)\citenamefont {Martin},
  \citenamefont {Papanikolaou}, \citenamefont {Pinol},\ and\ \citenamefont
  {Vennin}}]{Martin:2020fgl}%
  \BibitemOpen
  \bibfield  {author} {\bibinfo {author} {\bibfnamefont {J.}~\bibnamefont
  {Martin}}, \bibinfo {author} {\bibfnamefont {T.}~\bibnamefont
  {Papanikolaou}}, \bibinfo {author} {\bibfnamefont {L.}~\bibnamefont
  {Pinol}},\ and\ \bibinfo {author} {\bibfnamefont {V.}~\bibnamefont
  {Vennin}},\ }\href {https://doi.org/10.1088/1475-7516/2020/05/003} {\bibfield
   {journal} {\bibinfo  {journal} {JCAP}\ }\textbf {\bibinfo {volume} {05}},\
  \bibinfo {pages} {003}},\ \Eprint {https://arxiv.org/abs/2002.01820}
  {arXiv:2002.01820 [astro-ph.CO]} \BibitemShut {NoStop}%
\bibitem [{\citenamefont {Carr}(1975)}]{1975ApJ...201....1C}%
  \BibitemOpen
  \bibfield  {author} {\bibinfo {author} {\bibfnamefont {B.~J.}\ \bibnamefont
  {Carr}},\ }\href {https://doi.org/10.1086/153853} {\bibfield  {journal}
  {\bibinfo  {journal} {Astrophys. J.}\ }\textbf {\bibinfo {volume} {201}},\
  \bibinfo {pages} {1} (\bibinfo {year} {1975})}\BibitemShut {NoStop}%
\bibitem [{\citenamefont {Akrami}\ \emph {et~al.}(2020)\citenamefont {Akrami}
  \emph {et~al.}}]{Planck:2018jri}%
  \BibitemOpen
  \bibfield  {author} {\bibinfo {author} {\bibfnamefont {Y.}~\bibnamefont
  {Akrami}} \emph {et~al.} (\bibinfo {collaboration} {Planck}),\ }\href
  {https://doi.org/10.1051/0004-6361/201833887} {\bibfield  {journal} {\bibinfo
   {journal} {Astron. Astrophys.}\ }\textbf {\bibinfo {volume} {641}},\
  \bibinfo {pages} {A10} (\bibinfo {year} {2020})},\ \Eprint
  {https://arxiv.org/abs/1807.06211} {arXiv:1807.06211 [astro-ph.CO]}
  \BibitemShut {NoStop}%
\bibitem [{\citenamefont {Kawasaki}\ \emph {et~al.}(1999)\citenamefont
  {Kawasaki}, \citenamefont {Kohri},\ and\ \citenamefont
  {Sugiyama}}]{Kawasaki:1999na}%
  \BibitemOpen
  \bibfield  {author} {\bibinfo {author} {\bibfnamefont {M.}~\bibnamefont
  {Kawasaki}}, \bibinfo {author} {\bibfnamefont {K.}~\bibnamefont {Kohri}},\
  and\ \bibinfo {author} {\bibfnamefont {N.}~\bibnamefont {Sugiyama}},\ }\href
  {https://doi.org/10.1103/PhysRevLett.82.4168} {\bibfield  {journal} {\bibinfo
   {journal} {Phys. Rev. Lett.}\ }\textbf {\bibinfo {volume} {82}},\ \bibinfo
  {pages} {4168} (\bibinfo {year} {1999})},\ \Eprint
  {https://arxiv.org/abs/astro-ph/9811437} {arXiv:astro-ph/9811437}
  \BibitemShut {NoStop}%
\bibitem [{\citenamefont {Kawasaki}\ \emph {et~al.}(2000)\citenamefont
  {Kawasaki}, \citenamefont {Kohri},\ and\ \citenamefont
  {Sugiyama}}]{Kawasaki:2000en}%
  \BibitemOpen
  \bibfield  {author} {\bibinfo {author} {\bibfnamefont {M.}~\bibnamefont
  {Kawasaki}}, \bibinfo {author} {\bibfnamefont {K.}~\bibnamefont {Kohri}},\
  and\ \bibinfo {author} {\bibfnamefont {N.}~\bibnamefont {Sugiyama}},\ }\href
  {https://doi.org/10.1103/PhysRevD.62.023506} {\bibfield  {journal} {\bibinfo
  {journal} {Phys. Rev. D}\ }\textbf {\bibinfo {volume} {62}},\ \bibinfo
  {pages} {023506} (\bibinfo {year} {2000})},\ \Eprint
  {https://arxiv.org/abs/astro-ph/0002127} {arXiv:astro-ph/0002127}
  \BibitemShut {NoStop}%
\bibitem [{\citenamefont {Hannestad}(2004)}]{Hannestad:2004px}%
  \BibitemOpen
  \bibfield  {author} {\bibinfo {author} {\bibfnamefont {S.}~\bibnamefont
  {Hannestad}},\ }\href {https://doi.org/10.1103/PhysRevD.70.043506} {\bibfield
   {journal} {\bibinfo  {journal} {Phys. Rev. D}\ }\textbf {\bibinfo {volume}
  {70}},\ \bibinfo {pages} {043506} (\bibinfo {year} {2004})},\ \Eprint
  {https://arxiv.org/abs/astro-ph/0403291} {arXiv:astro-ph/0403291}
  \BibitemShut {NoStop}%
\bibitem [{\citenamefont {Hasegawa}\ \emph {et~al.}(2019)\citenamefont
  {Hasegawa}, \citenamefont {Hiroshima}, \citenamefont {Kohri}, \citenamefont
  {Hansen}, \citenamefont {Tram},\ and\ \citenamefont
  {Hannestad}}]{Hasegawa:2019jsa}%
  \BibitemOpen
  \bibfield  {author} {\bibinfo {author} {\bibfnamefont {T.}~\bibnamefont
  {Hasegawa}}, \bibinfo {author} {\bibfnamefont {N.}~\bibnamefont {Hiroshima}},
  \bibinfo {author} {\bibfnamefont {K.}~\bibnamefont {Kohri}}, \bibinfo
  {author} {\bibfnamefont {R.~S.~L.}\ \bibnamefont {Hansen}}, \bibinfo {author}
  {\bibfnamefont {T.}~\bibnamefont {Tram}},\ and\ \bibinfo {author}
  {\bibfnamefont {S.}~\bibnamefont {Hannestad}},\ }\href
  {https://doi.org/10.1088/1475-7516/2019/12/012} {\bibfield  {journal}
  {\bibinfo  {journal} {JCAP}\ }\textbf {\bibinfo {volume} {12}},\ \bibinfo
  {pages} {012}},\ \Eprint {https://arxiv.org/abs/1908.10189} {arXiv:1908.10189
  [hep-ph]} \BibitemShut {NoStop}%
\bibitem [{\citenamefont {Inomata}\ \emph
  {et~al.}(2019{\natexlab{a}})\citenamefont {Inomata}, \citenamefont {Kohri},
  \citenamefont {Nakama},\ and\ \citenamefont {Terada}}]{Inomata:2019ivs}%
  \BibitemOpen
  \bibfield  {author} {\bibinfo {author} {\bibfnamefont {K.}~\bibnamefont
  {Inomata}}, \bibinfo {author} {\bibfnamefont {K.}~\bibnamefont {Kohri}},
  \bibinfo {author} {\bibfnamefont {T.}~\bibnamefont {Nakama}},\ and\ \bibinfo
  {author} {\bibfnamefont {T.}~\bibnamefont {Terada}},\ }\href
  {https://doi.org/10.1103/PhysRevD.108.049901} {\bibfield  {journal} {\bibinfo
   {journal} {Phys. Rev. D}\ }\textbf {\bibinfo {volume} {100}},\ \bibinfo
  {pages} {043532} (\bibinfo {year} {2019}{\natexlab{a}})},\ \bibinfo {note}
  {[Erratum: Phys.Rev.D 108, 049901 (2023)]},\ \Eprint
  {https://arxiv.org/abs/1904.12879} {arXiv:1904.12879 [astro-ph.CO]}
  \BibitemShut {NoStop}%
\bibitem [{\citenamefont {Inomata}\ \emph
  {et~al.}(2019{\natexlab{b}})\citenamefont {Inomata}, \citenamefont {Kohri},
  \citenamefont {Nakama},\ and\ \citenamefont {Terada}}]{Inomata:2019zqy}%
  \BibitemOpen
  \bibfield  {author} {\bibinfo {author} {\bibfnamefont {K.}~\bibnamefont
  {Inomata}}, \bibinfo {author} {\bibfnamefont {K.}~\bibnamefont {Kohri}},
  \bibinfo {author} {\bibfnamefont {T.}~\bibnamefont {Nakama}},\ and\ \bibinfo
  {author} {\bibfnamefont {T.}~\bibnamefont {Terada}},\ }\href
  {https://doi.org/10.1088/1475-7516/2019/10/071} {\bibfield  {journal}
  {\bibinfo  {journal} {JCAP}\ }\textbf {\bibinfo {volume} {10}},\ \bibinfo
  {pages} {071}},\ \bibinfo {note} {[Erratum: JCAP 08, E01 (2023)]},\ \Eprint
  {https://arxiv.org/abs/1904.12878} {arXiv:1904.12878 [astro-ph.CO]}
  \BibitemShut {NoStop}%
\bibitem [{\citenamefont {Dom\`enech}\ and\ \citenamefont
  {Sasaki}(2024)}]{Domenech:2024cjn}%
  \BibitemOpen
  \bibfield  {author} {\bibinfo {author} {\bibfnamefont {G.}~\bibnamefont
  {Dom\`enech}}\ and\ \bibinfo {author} {\bibfnamefont {M.}~\bibnamefont
  {Sasaki}},\ }\href@noop {} {\  (\bibinfo {year} {2024})},\ \Eprint
  {https://arxiv.org/abs/2401.07615} {arXiv:2401.07615 [gr-qc]} \BibitemShut
  {NoStop}%
\bibitem [{\citenamefont {Dom\`enech}(2024{\natexlab{b}})}]{Domenech:2024kmh}%
  \BibitemOpen
  \bibfield  {author} {\bibinfo {author} {\bibfnamefont {G.}~\bibnamefont
  {Dom\`enech}},\ }\href@noop {} {\  (\bibinfo {year} {2024}{\natexlab{b}})},\
  \Eprint {https://arxiv.org/abs/2402.17388} {arXiv:2402.17388 [gr-qc]}
  \BibitemShut {NoStop}%
\bibitem [{\citenamefont {Thrane}\ and\ \citenamefont
  {Romano}(2013)}]{Thrane:2013oya}%
  \BibitemOpen
  \bibfield  {author} {\bibinfo {author} {\bibfnamefont {E.}~\bibnamefont
  {Thrane}}\ and\ \bibinfo {author} {\bibfnamefont {J.~D.}\ \bibnamefont
  {Romano}},\ }\href {https://doi.org/10.1103/PhysRevD.88.124032} {\bibfield
  {journal} {\bibinfo  {journal} {Phys. Rev.}\ }\textbf {\bibinfo {volume}
  {D88}},\ \bibinfo {pages} {124032} (\bibinfo {year} {2013})},\ \Eprint
  {https://arxiv.org/abs/1310.5300} {arXiv:1310.5300 [astro-ph.IM]}
  \BibitemShut {NoStop}%
\bibitem [{\citenamefont {Schmitz}(2021)}]{Schmitz:2020syl}%
  \BibitemOpen
  \bibfield  {author} {\bibinfo {author} {\bibfnamefont {K.}~\bibnamefont
  {Schmitz}},\ }\href {https://doi.org/10.1007/JHEP01(2021)097} {\bibfield
  {journal} {\bibinfo  {journal} {JHEP}\ }\textbf {\bibinfo {volume} {01}},\
  \bibinfo {pages} {097}},\ \Eprint {https://arxiv.org/abs/2002.04615}
  {arXiv:2002.04615 [hep-ph]} \BibitemShut {NoStop}%
\bibitem [{\citenamefont {Branchesi}\ \emph {et~al.}(2023)\citenamefont
  {Branchesi} \emph {et~al.}}]{Branchesi:2023mws}%
  \BibitemOpen
  \bibfield  {author} {\bibinfo {author} {\bibfnamefont {M.}~\bibnamefont
  {Branchesi}} \emph {et~al.},\ }\href
  {https://doi.org/10.1088/1475-7516/2023/07/068} {\bibfield  {journal}
  {\bibinfo  {journal} {JCAP}\ }\textbf {\bibinfo {volume} {07}},\ \bibinfo
  {pages} {068}},\ \Eprint {https://arxiv.org/abs/2303.15923} {arXiv:2303.15923
  [gr-qc]} \BibitemShut {NoStop}%
\bibitem [{ce()}]{ce}%
  \BibitemOpen
  \href@noop {} {\bibinfo {title} {Cosmic explorer sensitivity curve}},\
  \bibinfo {howpublished} {\url{https://cosmicexplorer.org/sensitivity.html}},\
  \bibinfo {note} {[Online; accessed 05-May-2023]}\BibitemShut {NoStop}%
\bibitem [{A+()}]{A+}%
  \BibitemOpen
  \href@noop {} {\bibinfo {title} {The {A+} design curve}},\ \bibinfo
  {howpublished} {\url{https://dcc.ligo.org/LIGO-T1800042/public}},\ \bibinfo
  {note} {[Online; accessed 05-May-2023]}\BibitemShut {NoStop}%
\bibitem [{voy()}]{voyager}%
  \BibitemOpen
  \href@noop {} {\bibinfo {title} {Ligo unofficial sensitivity curves}},\
  \bibinfo {howpublished} {\url{https://dcc.ligo.org/LIGO-T1500293/public}},\
  \bibinfo {note} {[Online; accessed 05-May-2023]}\BibitemShut {NoStop}%
\bibitem [{\citenamefont {Yagi}\ and\ \citenamefont
  {Seto}(2011)}]{Yagi:2011wg}%
  \BibitemOpen
  \bibfield  {author} {\bibinfo {author} {\bibfnamefont {K.}~\bibnamefont
  {Yagi}}\ and\ \bibinfo {author} {\bibfnamefont {N.}~\bibnamefont {Seto}},\
  }\href {https://doi.org/10.1103/PhysRevD.95.109901,
  10.1103/PhysRevD.83.044011} {\bibfield  {journal} {\bibinfo  {journal} {Phys.
  Rev.}\ }\textbf {\bibinfo {volume} {D83}},\ \bibinfo {pages} {044011}
  (\bibinfo {year} {2011})},\ \bibinfo {note} {[Erratum: Phys.
  Rev.D95,no.10,109901(2017)]},\ \Eprint {https://arxiv.org/abs/1101.3940}
  {arXiv:1101.3940 [astro-ph.CO]} \BibitemShut {NoStop}%
\bibitem [{\citenamefont {Kawamura}\ \emph {et~al.}(2020)\citenamefont
  {Kawamura} \emph {et~al.}}]{Kawamura:2020pcg}%
  \BibitemOpen
  \bibfield  {author} {\bibinfo {author} {\bibfnamefont {S.}~\bibnamefont
  {Kawamura}} \emph {et~al.},\ }\href@noop {} {\  (\bibinfo {year} {2020})},\
  \Eprint {https://arxiv.org/abs/2006.13545} {arXiv:2006.13545 [gr-qc]}
  \BibitemShut {NoStop}%
\bibitem [{\citenamefont {Abbott}\ \emph {et~al.}(2021)\citenamefont {Abbott}
  \emph {et~al.}}]{KAGRA:2021kbb}%
  \BibitemOpen
  \bibfield  {author} {\bibinfo {author} {\bibfnamefont {R.}~\bibnamefont
  {Abbott}} \emph {et~al.} (\bibinfo {collaboration} {KAGRA, Virgo, LIGO
  Scientific}),\ }\href {https://doi.org/10.1103/PhysRevD.104.022004}
  {\bibfield  {journal} {\bibinfo  {journal} {Phys. Rev. D}\ }\textbf {\bibinfo
  {volume} {104}},\ \bibinfo {pages} {022004} (\bibinfo {year} {2021})},\
  \Eprint {https://arxiv.org/abs/2101.12130} {arXiv:2101.12130 [gr-qc]}
  \BibitemShut {NoStop}%
\bibitem [{\citenamefont {Cyburt}\ \emph {et~al.}(2005)\citenamefont {Cyburt},
  \citenamefont {Fields}, \citenamefont {Olive},\ and\ \citenamefont
  {Skillman}}]{Cyburt:2004yc}%
  \BibitemOpen
  \bibfield  {author} {\bibinfo {author} {\bibfnamefont {R.~H.}\ \bibnamefont
  {Cyburt}}, \bibinfo {author} {\bibfnamefont {B.~D.}\ \bibnamefont {Fields}},
  \bibinfo {author} {\bibfnamefont {K.~A.}\ \bibnamefont {Olive}},\ and\
  \bibinfo {author} {\bibfnamefont {E.}~\bibnamefont {Skillman}},\ }\href
  {https://doi.org/10.1016/j.astropartphys.2005.01.005} {\bibfield  {journal}
  {\bibinfo  {journal} {Astropart. Phys.}\ }\textbf {\bibinfo {volume} {23}},\
  \bibinfo {pages} {313} (\bibinfo {year} {2005})},\ \Eprint
  {https://arxiv.org/abs/astro-ph/0408033} {arXiv:astro-ph/0408033}
  \BibitemShut {NoStop}%
\bibitem [{\citenamefont {Arbey}\ \emph {et~al.}(2021)\citenamefont {Arbey},
  \citenamefont {Auffinger}, \citenamefont {Sandick}, \citenamefont {Shams
  Es~Haghi},\ and\ \citenamefont {Sinha}}]{Arbey:2021ysg}%
  \BibitemOpen
  \bibfield  {author} {\bibinfo {author} {\bibfnamefont {A.}~\bibnamefont
  {Arbey}}, \bibinfo {author} {\bibfnamefont {J.}~\bibnamefont {Auffinger}},
  \bibinfo {author} {\bibfnamefont {P.}~\bibnamefont {Sandick}}, \bibinfo
  {author} {\bibfnamefont {B.}~\bibnamefont {Shams Es~Haghi}},\ and\ \bibinfo
  {author} {\bibfnamefont {K.}~\bibnamefont {Sinha}},\ }\href
  {https://doi.org/10.1103/PhysRevD.103.123549} {\bibfield  {journal} {\bibinfo
   {journal} {Phys. Rev. D}\ }\textbf {\bibinfo {volume} {103}},\ \bibinfo
  {pages} {123549} (\bibinfo {year} {2021})},\ \Eprint
  {https://arxiv.org/abs/2104.04051} {arXiv:2104.04051 [astro-ph.CO]}
  \BibitemShut {NoStop}%
\bibitem [{\citenamefont {Grohs}\ and\ \citenamefont
  {Fuller}(2023)}]{Grohs:2023voo}%
  \BibitemOpen
  \bibfield  {author} {\bibinfo {author} {\bibfnamefont {E.}~\bibnamefont
  {Grohs}}\ and\ \bibinfo {author} {\bibfnamefont {G.~M.}\ \bibnamefont
  {Fuller}},\ }\bibinfo {title} {{Big Bang Nucleosynthesis}},\ in\ \href
  {https://doi.org/10.1007/978-981-15-8818-1_127-1} {\emph {\bibinfo
  {booktitle} {{Handbook of Nuclear Physics}}}},\ \bibinfo {editor} {edited by\
  \bibinfo {editor} {\bibfnamefont {I.}~\bibnamefont {Tanihata}}, \bibinfo
  {editor} {\bibfnamefont {H.}~\bibnamefont {Toki}},\ and\ \bibinfo {editor}
  {\bibfnamefont {T.}~\bibnamefont {Kajino}}}\ (\bibinfo {year} {2023})\ pp.\
  \bibinfo {pages} {1--21},\ \Eprint {https://arxiv.org/abs/2301.12299}
  {arXiv:2301.12299 [astro-ph.CO]} \BibitemShut {NoStop}%
\bibitem [{\citenamefont {Caprini}\ and\ \citenamefont
  {Figueroa}(2018)}]{Caprini:2018mtu}%
  \BibitemOpen
  \bibfield  {author} {\bibinfo {author} {\bibfnamefont {C.}~\bibnamefont
  {Caprini}}\ and\ \bibinfo {author} {\bibfnamefont {D.~G.}\ \bibnamefont
  {Figueroa}},\ }\href {https://doi.org/10.1088/1361-6382/aac608} {\bibfield
  {journal} {\bibinfo  {journal} {Class. Quant. Grav.}\ }\textbf {\bibinfo
  {volume} {35}},\ \bibinfo {pages} {163001} (\bibinfo {year} {2018})},\
  \Eprint {https://arxiv.org/abs/1801.04268} {arXiv:1801.04268 [astro-ph.CO]}
  \BibitemShut {NoStop}%
\bibitem [{\citenamefont {Dong}\ \emph {et~al.}(2016)\citenamefont {Dong},
  \citenamefont {Kinney},\ and\ \citenamefont {Stojkovic}}]{Dong:2015yjs}%
  \BibitemOpen
  \bibfield  {author} {\bibinfo {author} {\bibfnamefont {R.}~\bibnamefont
  {Dong}}, \bibinfo {author} {\bibfnamefont {W.~H.}\ \bibnamefont {Kinney}},\
  and\ \bibinfo {author} {\bibfnamefont {D.}~\bibnamefont {Stojkovic}},\ }\href
  {https://doi.org/10.1088/1475-7516/2016/10/034} {\bibfield  {journal}
  {\bibinfo  {journal} {JCAP}\ }\textbf {\bibinfo {volume} {10}},\ \bibinfo
  {pages} {034}},\ \Eprint {https://arxiv.org/abs/1511.05642} {arXiv:1511.05642
  [astro-ph.CO]} \BibitemShut {NoStop}%
\bibitem [{\citenamefont {Arbey}\ \emph {et~al.}(2020)\citenamefont {Arbey},
  \citenamefont {Auffinger},\ and\ \citenamefont {Silk}}]{Arbey:2019jmj}%
  \BibitemOpen
  \bibfield  {author} {\bibinfo {author} {\bibfnamefont {A.}~\bibnamefont
  {Arbey}}, \bibinfo {author} {\bibfnamefont {J.}~\bibnamefont {Auffinger}},\
  and\ \bibinfo {author} {\bibfnamefont {J.}~\bibnamefont {Silk}},\ }\href
  {https://doi.org/10.1093/mnras/staa765} {\bibfield  {journal} {\bibinfo
  {journal} {Mon. Not. Roy. Astron. Soc.}\ }\textbf {\bibinfo {volume} {494}},\
  \bibinfo {pages} {1257} (\bibinfo {year} {2020})},\ \Eprint
  {https://arxiv.org/abs/1906.04196} {arXiv:1906.04196 [astro-ph.CO]}
  \BibitemShut {NoStop}%
\bibitem [{\citenamefont {MacGibbon}(1987)}]{MacGibbon:1987my}%
  \BibitemOpen
  \bibfield  {author} {\bibinfo {author} {\bibfnamefont {J.~H.}\ \bibnamefont
  {MacGibbon}},\ }\href {https://doi.org/10.1038/329308a0} {\bibfield
  {journal} {\bibinfo  {journal} {Nature}\ }\textbf {\bibinfo {volume} {329}},\
  \bibinfo {pages} {308} (\bibinfo {year} {1987})}\BibitemShut {NoStop}%
\bibitem [{\citenamefont {Chen}\ \emph {et~al.}(2015)\citenamefont {Chen},
  \citenamefont {Ong},\ and\ \citenamefont {Yeom}}]{Chen:2014jwq}%
  \BibitemOpen
  \bibfield  {author} {\bibinfo {author} {\bibfnamefont {P.}~\bibnamefont
  {Chen}}, \bibinfo {author} {\bibfnamefont {Y.~C.}\ \bibnamefont {Ong}},\ and\
  \bibinfo {author} {\bibfnamefont {D.-h.}\ \bibnamefont {Yeom}},\ }\href
  {https://doi.org/10.1016/j.physrep.2015.10.007} {\bibfield  {journal}
  {\bibinfo  {journal} {Phys. Rept.}\ }\textbf {\bibinfo {volume} {603}},\
  \bibinfo {pages} {1} (\bibinfo {year} {2015})},\ \Eprint
  {https://arxiv.org/abs/1412.8366} {arXiv:1412.8366 [gr-qc]} \BibitemShut
  {NoStop}%
\bibitem [{\citenamefont {Hossenfelder}(2013)}]{Hossenfelder:2012jw}%
  \BibitemOpen
  \bibfield  {author} {\bibinfo {author} {\bibfnamefont {S.}~\bibnamefont
  {Hossenfelder}},\ }\href {https://doi.org/10.12942/lrr-2013-2} {\bibfield
  {journal} {\bibinfo  {journal} {Living Rev. Rel.}\ }\textbf {\bibinfo
  {volume} {16}},\ \bibinfo {pages} {2} (\bibinfo {year} {2013})},\ \Eprint
  {https://arxiv.org/abs/1203.6191} {arXiv:1203.6191 [gr-qc]} \BibitemShut
  {NoStop}%
\bibitem [{\citenamefont {Vidotto}(2018)}]{Vidotto:2018hww}%
  \BibitemOpen
  \bibfield  {author} {\bibinfo {author} {\bibfnamefont {F.}~\bibnamefont
  {Vidotto}},\ }\href {https://doi.org/10.22323/1.335.0046} {\bibfield
  {journal} {\bibinfo  {journal} {PoS}\ }\textbf {\bibinfo {volume}
  {EDSU2018}},\ \bibinfo {pages} {046} (\bibinfo {year} {2018})},\ \Eprint
  {https://arxiv.org/abs/1811.08007} {arXiv:1811.08007 [gr-qc]} \BibitemShut
  {NoStop}%
\bibitem [{\citenamefont {Eichhorn}\ and\ \citenamefont
  {Held}(2022)}]{Eichhorn:2022bgu}%
  \BibitemOpen
  \bibfield  {author} {\bibinfo {author} {\bibfnamefont {A.}~\bibnamefont
  {Eichhorn}}\ and\ \bibinfo {author} {\bibfnamefont {A.}~\bibnamefont
  {Held}},\ }\href@noop {} {\  (\bibinfo {year} {2022})},\ \Eprint
  {https://arxiv.org/abs/2212.09495} {arXiv:2212.09495 [gr-qc]} \BibitemShut
  {NoStop}%
\bibitem [{\citenamefont {Platania}(2023)}]{Platania:2023srt}%
  \BibitemOpen
  \bibfield  {author} {\bibinfo {author} {\bibfnamefont {A.}~\bibnamefont
  {Platania}},\ }\bibinfo {title} {{Black Holes in Asymptotically Safe
  Gravity}}\ (\bibinfo {year} {2023})\ \Eprint
  {https://arxiv.org/abs/2302.04272} {arXiv:2302.04272 [gr-qc]} \BibitemShut
  {NoStop}%
\bibitem [{\citenamefont {Dom\`enech}\ and\ \citenamefont
  {Sasaki}(2023)}]{Domenech:2023mqk}%
  \BibitemOpen
  \bibfield  {author} {\bibinfo {author} {\bibfnamefont {G.}~\bibnamefont
  {Dom\`enech}}\ and\ \bibinfo {author} {\bibfnamefont {M.}~\bibnamefont
  {Sasaki}},\ }\href {https://doi.org/10.1088/1361-6382/ace493} {\bibfield
  {journal} {\bibinfo  {journal} {Class. Quant. Grav.}\ }\textbf {\bibinfo
  {volume} {40}},\ \bibinfo {pages} {177001} (\bibinfo {year} {2023})},\
  \Eprint {https://arxiv.org/abs/2303.07661} {arXiv:2303.07661 [gr-qc]}
  \BibitemShut {NoStop}%
\bibitem [{\citenamefont {Papanikolaou}\ \emph {et~al.}(2024)\citenamefont
  {Papanikolaou}, \citenamefont {He}, \citenamefont {Ma}, \citenamefont {Cai},
  \citenamefont {Saridakis},\ and\ \citenamefont
  {Sasaki}}]{Papanikolaou:2024kjb}%
  \BibitemOpen
  \bibfield  {author} {\bibinfo {author} {\bibfnamefont {T.}~\bibnamefont
  {Papanikolaou}}, \bibinfo {author} {\bibfnamefont {X.-C.}\ \bibnamefont
  {He}}, \bibinfo {author} {\bibfnamefont {X.-H.}\ \bibnamefont {Ma}}, \bibinfo
  {author} {\bibfnamefont {Y.-F.}\ \bibnamefont {Cai}}, \bibinfo {author}
  {\bibfnamefont {E.~N.}\ \bibnamefont {Saridakis}},\ and\ \bibinfo {author}
  {\bibfnamefont {M.}~\bibnamefont {Sasaki}},\ }\href
  {https://doi.org/10.1016/j.physletb.2024.138997} {\bibfield  {journal}
  {\bibinfo  {journal} {Phys. Lett. B}\ }\textbf {\bibinfo {volume} {857}},\
  \bibinfo {pages} {138997} (\bibinfo {year} {2024})},\ \Eprint
  {https://arxiv.org/abs/2403.00660} {arXiv:2403.00660 [astro-ph.CO]}
  \BibitemShut {NoStop}%
\bibitem [{\citenamefont {He}\ \emph {et~al.}(2024)\citenamefont {He},
  \citenamefont {Cai}, \citenamefont {Ma}, \citenamefont {Papanikolaou},
  \citenamefont {Saridakis},\ and\ \citenamefont {Sasaki}}]{He:2024luf}%
  \BibitemOpen
  \bibfield  {author} {\bibinfo {author} {\bibfnamefont {X.-C.}\ \bibnamefont
  {He}}, \bibinfo {author} {\bibfnamefont {Y.-F.}\ \bibnamefont {Cai}},
  \bibinfo {author} {\bibfnamefont {X.-H.}\ \bibnamefont {Ma}}, \bibinfo
  {author} {\bibfnamefont {T.}~\bibnamefont {Papanikolaou}}, \bibinfo {author}
  {\bibfnamefont {E.~N.}\ \bibnamefont {Saridakis}},\ and\ \bibinfo {author}
  {\bibfnamefont {M.}~\bibnamefont {Sasaki}},\ }\href@noop {} {\  (\bibinfo
  {year} {2024})},\ \Eprint {https://arxiv.org/abs/2409.11333}
  {arXiv:2409.11333 [astro-ph.CO]} \BibitemShut {NoStop}%
\bibitem [{\citenamefont {Briscese}\ and\ \citenamefont
  {Elizalde}(2008)}]{Briscese:2007cd}%
  \BibitemOpen
  \bibfield  {author} {\bibinfo {author} {\bibfnamefont {F.}~\bibnamefont
  {Briscese}}\ and\ \bibinfo {author} {\bibfnamefont {E.}~\bibnamefont
  {Elizalde}},\ }\href {https://doi.org/10.1103/PhysRevD.77.044009} {\bibfield
  {journal} {\bibinfo  {journal} {Phys. Rev. D}\ }\textbf {\bibinfo {volume}
  {77}},\ \bibinfo {pages} {044009} (\bibinfo {year} {2008})},\ \Eprint
  {https://arxiv.org/abs/0708.0432} {arXiv:0708.0432 [hep-th]} \BibitemShut
  {NoStop}%
\bibitem [{\citenamefont {Mureika}\ \emph {et~al.}(2016)\citenamefont
  {Mureika}, \citenamefont {Moffat},\ and\ \citenamefont
  {Faizal}}]{Mureika:2015sda}%
  \BibitemOpen
  \bibfield  {author} {\bibinfo {author} {\bibfnamefont {J.~R.}\ \bibnamefont
  {Mureika}}, \bibinfo {author} {\bibfnamefont {J.~W.}\ \bibnamefont
  {Moffat}},\ and\ \bibinfo {author} {\bibfnamefont {M.}~\bibnamefont
  {Faizal}},\ }\href {https://doi.org/10.1016/j.physletb.2016.04.041}
  {\bibfield  {journal} {\bibinfo  {journal} {Phys. Lett. B}\ }\textbf
  {\bibinfo {volume} {757}},\ \bibinfo {pages} {528} (\bibinfo {year}
  {2016})},\ \Eprint {https://arxiv.org/abs/1504.08226} {arXiv:1504.08226
  [gr-qc]} \BibitemShut {NoStop}%
\bibitem [{\citenamefont {Calmet}\ and\ \citenamefont
  {Kuipers}(2021)}]{Calmet:2021lny}%
  \BibitemOpen
  \bibfield  {author} {\bibinfo {author} {\bibfnamefont {X.}~\bibnamefont
  {Calmet}}\ and\ \bibinfo {author} {\bibfnamefont {F.}~\bibnamefont
  {Kuipers}},\ }\href {https://doi.org/10.1103/PhysRevD.104.066012} {\bibfield
  {journal} {\bibinfo  {journal} {Phys. Rev. D}\ }\textbf {\bibinfo {volume}
  {104}},\ \bibinfo {pages} {066012} (\bibinfo {year} {2021})},\ \Eprint
  {https://arxiv.org/abs/2108.06824} {arXiv:2108.06824 [hep-th]} \BibitemShut
  {NoStop}%
\bibitem [{\citenamefont {Dolgov}\ and\ \citenamefont
  {Ejlli}(2011)}]{Dolgov:2011cq}%
  \BibitemOpen
  \bibfield  {author} {\bibinfo {author} {\bibfnamefont {A.~D.}\ \bibnamefont
  {Dolgov}}\ and\ \bibinfo {author} {\bibfnamefont {D.}~\bibnamefont {Ejlli}},\
  }\href {https://doi.org/10.1103/PhysRevD.84.024028} {\bibfield  {journal}
  {\bibinfo  {journal} {Phys. Rev. D}\ }\textbf {\bibinfo {volume} {84}},\
  \bibinfo {pages} {024028} (\bibinfo {year} {2011})},\ \Eprint
  {https://arxiv.org/abs/1105.2303} {arXiv:1105.2303 [astro-ph.CO]}
  \BibitemShut {NoStop}%
\bibitem [{\citenamefont {Masina}(2021)}]{Masina:2021zpu}%
  \BibitemOpen
  \bibfield  {author} {\bibinfo {author} {\bibfnamefont {I.}~\bibnamefont
  {Masina}},\ }\href {https://doi.org/10.1134/S0202289321040101} {\bibfield
  {journal} {\bibinfo  {journal} {Grav. Cosmol.}\ }\textbf {\bibinfo {volume}
  {27}},\ \bibinfo {pages} {315} (\bibinfo {year} {2021})},\ \Eprint
  {https://arxiv.org/abs/2103.13825} {arXiv:2103.13825 [gr-qc]} \BibitemShut
  {NoStop}%
\bibitem [{\citenamefont {Gehrman}\ \emph {et~al.}(2023)\citenamefont
  {Gehrman}, \citenamefont {Shams Es~Haghi}, \citenamefont {Sinha},\ and\
  \citenamefont {Xu}}]{Gehrman:2023esa}%
  \BibitemOpen
  \bibfield  {author} {\bibinfo {author} {\bibfnamefont {T.~C.}\ \bibnamefont
  {Gehrman}}, \bibinfo {author} {\bibfnamefont {B.}~\bibnamefont {Shams
  Es~Haghi}}, \bibinfo {author} {\bibfnamefont {K.}~\bibnamefont {Sinha}},\
  and\ \bibinfo {author} {\bibfnamefont {T.}~\bibnamefont {Xu}},\ }\href
  {https://doi.org/10.1088/1475-7516/2023/10/001} {\bibfield  {journal}
  {\bibinfo  {journal} {JCAP}\ }\textbf {\bibinfo {volume} {10}},\ \bibinfo
  {pages} {001}},\ \Eprint {https://arxiv.org/abs/2304.09194} {arXiv:2304.09194
  [hep-ph]} \BibitemShut {NoStop}%
\bibitem [{\citenamefont {Ireland}\ \emph {et~al.}(2023)\citenamefont
  {Ireland}, \citenamefont {Profumo},\ and\ \citenamefont
  {Scharnhorst}}]{Ireland:2023avg}%
  \BibitemOpen
  \bibfield  {author} {\bibinfo {author} {\bibfnamefont {A.}~\bibnamefont
  {Ireland}}, \bibinfo {author} {\bibfnamefont {S.}~\bibnamefont {Profumo}},\
  and\ \bibinfo {author} {\bibfnamefont {J.}~\bibnamefont {Scharnhorst}},\
  }\href {https://doi.org/10.1103/PhysRevD.107.104021} {\bibfield  {journal}
  {\bibinfo  {journal} {Phys. Rev. D}\ }\textbf {\bibinfo {volume} {107}},\
  \bibinfo {pages} {104021} (\bibinfo {year} {2023})},\ \Eprint
  {https://arxiv.org/abs/2302.10188} {arXiv:2302.10188 [gr-qc]} \BibitemShut
  {NoStop}%
\end{thebibliography}%
\let\addcontentsline\oldaddcontentsline
\end{document}